\documentclass[a4paper,fleqn,usenatbib]{mnras}

\usepackage[T1]{fontenc}
\usepackage{ae,aecompl}

\usepackage{graphicx}	% Including figure files
\usepackage{amsmath}	% Advanced maths commands
\usepackage{amssymb}	% Extra maths symbols
\usepackage{threeparttable}
\usepackage{tablefootnote}
\title[Central Parsec of M81]{Inflowing Gas in the Central Parsec of M81}

\author[Devereux]{
Nick Devereux$^{1}$\thanks{E-mail: devereux@erau.edu}
\\
$^{1}$Embry-Riddle Aeronautical University, 3700 Willow Creek Road, Prescott, AZ 8301, USA\\
}

\date{Accepted XXX. Received YYY; in original form ZZZ}

\pubyear{2018}

\begin{document}
\label{firstpage}
\pagerange{\pageref{firstpage}--\pageref{lastpage}}
\maketitle

\begin{abstract}

Spectroscopic observations of the Seyfert 1/Liner nucleus of M81, obtained recently with the 
Space Telescope Imaging Spectrograph (STIS) aboard the {\it Hubble Space Telescope (HST)}, have revealed a UV--visible spectrum rich with emission lines of a 
variety of widths, ionization potentials, and critical densities, including several in the UV that have not previously been reported. Even at the highest angular 
resolution currently achievable with {\it HST}, the broad-line region of M81 cannot be uniquely 
defined on the basis of commonly used observables such as the full-width at half maximum of the emission lines, or 
ratios of various emission lines. Numerous broad forbidden lines complicate interpretation of the spectra. 
At least three separate line-emitting components are inferred.  
A large, highly ionized, low density, low metallicity H${^+}$ region producing the broad Balmer lines. 
Located within the H${^+}$ region are smaller condensations spanning a wide-range in density, and the source of forbidden line emission through 
collisional excitation of the respective ions. Intermingled with the H${^+}$ region and the condensations
is a curious extended source of time-variable \ion{C}{iv} ${\lambda}$ 1548 emission. 
Collectively, these observations can be qualitatively understood in the context of
a shock excited jet cavity within a large H${^+}$ region that is photoionized by the central UV--X-ray source. The H${^+}$ region contains ${\sim}$ 500 M${\odot}$ of 
low metallicity gas that is dynamically unstable to inflow. At the current rate, the available H${^+}$ gas can sustain the advection dominated accretion flow that powers 
the central UV--X-ray source for 10$^{5}$ years.
 
\end{abstract}


%%%%%%%%%%%%%%%%%%%%%%%%%%%%%%%%%%%%%%%%%%%%%%%%%%
%Figures are referred to as e.g. Fig.~\ref{fig:example_figure}, and tables as
%e.g. Table~\ref{tab:example_table}.
%%%%%%%%%%%%%%%%% BODY OF PAPER %%%%%%%%%%%%%%%%%%

\section{Introduction}

M81, by virtue of its proximity, 3.6 Mpc \citep{Freedman2001a}, hosts the brightest and best resolved 
(17.5 pc arcsec$^{-1}$) active nucleus, albeit a very low luminosity one \citep{Petre1993a}. Classified as both Seyfert 1 \citep{Peimbert1981} and 
Liner \citep{Heckman1980}, the activity manifests as a broad-line region (BLR) centred in an extended region of shocked-ionized gas.
Broad emission lines appearing in visible and UV spectra were first described by \citet{Peimbert1981} and \citet{Bruzual1982}. They reasoned that the anomalous 
H${\alpha}$/H${\beta}$ ratio was the result of photoionization in clouds of high density,  ${\sim}$10$^9$ cm$^{-3}$,
and consequently, that the BLR in M81 must be very small, ${\le}$ 10$^{-2}$ pc in radius.
These ideas were reinforced by \citet{Filippenko1988} and \citet{Ho1996} who estimated a BLR size yet a further order of magnitude smaller. Significantly, neither 
the broad Balmer emission lines, nor the adjacent continuum
are time variable precluding an estimate of the BLR size using reverberation mapping \citep{Peterson2004, Kaspi2005}.
Nevertheless, should M81 conform to the BLR size-luminosity relationship established for more luminous reverberating Seyferts, the extraordinarily low 
luminosity measured for the Seyfert nucleus in M81, corresponding to 1.8 ${\times}$ 10$^{40}$ erg s$^{-1}$ at 5100 \AA, would imply that the BLR should be an 
order of magnitude smaller still, an implausible 10$^{-4}$ pc in radius, equivalent to about one-tenth of a light-day, comparable to the distance between the Sun 
and Uranus. This would cause the H${\alpha}$ line width to be far broader than is actually observed, as noted previously by \citet{Laor2003}, indicating that M81 
does not, in fact, conform to the BLR size-luminosity correlation established for reverberating AGN.

\begin{figure*}
	\includegraphics[width=\textwidth]{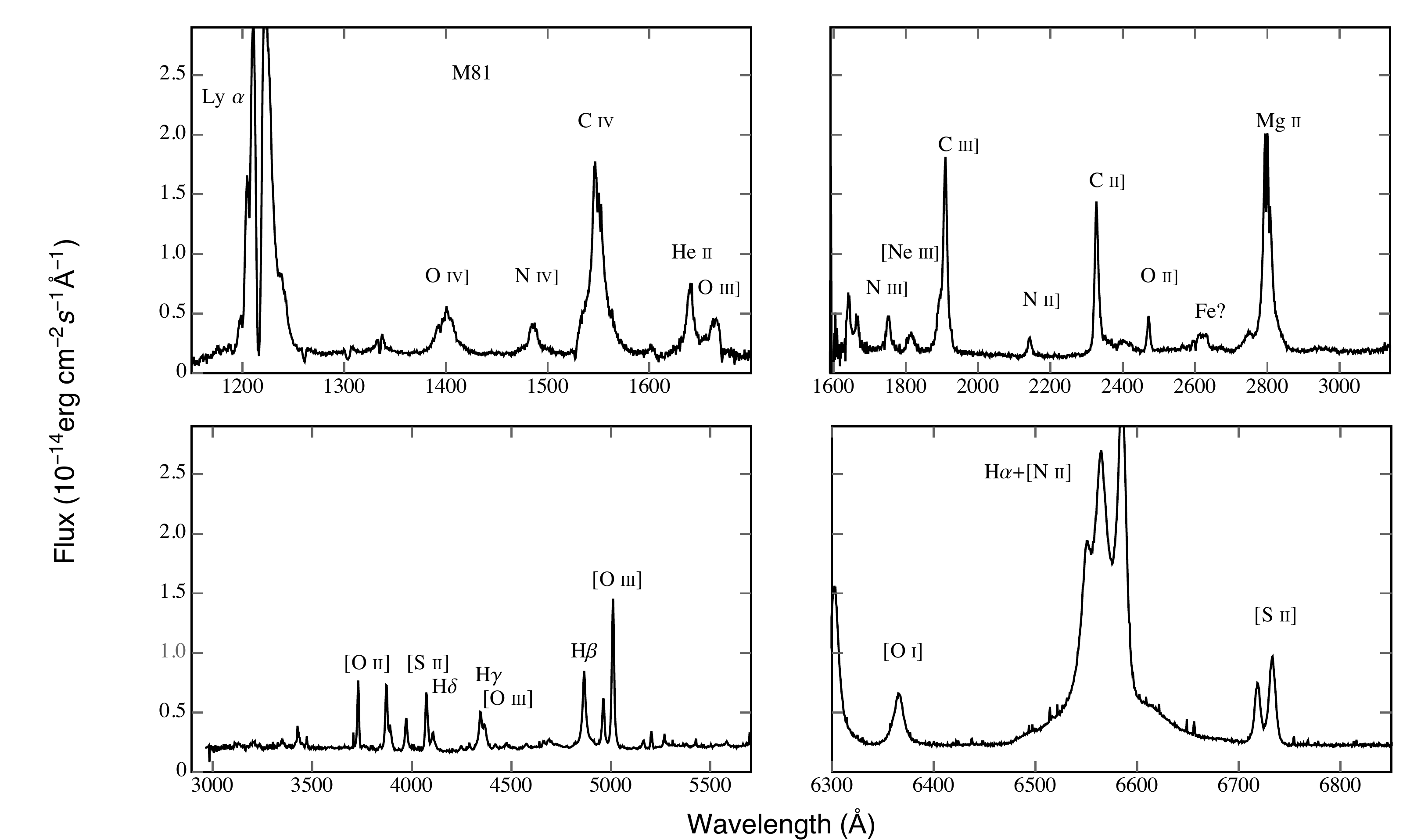}
	\caption{Visual and UV spectra  of M81 as seen through the following gratings:  {\sl Top left panel}: G140L. {\sl Top right panel}: G230L.
{\sl Lower left panel}: G430L. {\sl Lower right panel}: G750M.  The ordinate measures flux in units of erg cm$^{-2}$ s$^{-1}$ \AA$^{-1}$ 
whereas the abscissa is in units of {\AA}. }
    \label{fig:example_figure}
\end{figure*}

\begin{table*}
\centering
\caption{M81 STIS Datasets}
\begin{tabular}{lccccl}
\hline
Grating & Spectral Range & Dispersion & Plate Scale & Integration Time & Datasets   \\
& \AA & \AA/pix & \arcsec/pix & sec & \\
\hline
G750M & 6295-6867 & 0.56 & 0.05078 & 475 ${\times}$ 5 & odk803010 - odk803050 \\
G430L & 2900-5700 & 2.73 & 0.05078 & 368 ${\times}$ 5 & odk803060 - odk803090, odk8030A0 \\
G230L & 1570-3180 & 1.58 & 0.025 & 1035 ${\times}$ 3 & odk802040 - odk802060 \\
G140L & 1150-1730 & 0.6 & 0.0246 & 838 ${\times}$ 6 & odk801010-odk801030, odk802010 - odk802030 \\ 
G140L & 1150-1730 & 0.6 & 0.0246 &1066 ${\times}$ 6  &  odk801040-odk801090 \\
\hline
\end{tabular}
\end{table*}

M81 was one of the first galaxies targeted with the {\it Hubble Space Telescope} ({\it HST}) to yield a black hole (BH) mass, M$_{\bullet}$, using spatially resolved gas kinematics. Those observations suggest M$_{\bullet}$ = 7 ${^{+2}_{-1}}$ ${\times}$ 10$^{7}$ M${_{\sun}}$ \citep{Devereux2003}. Knowing the BH mass provides a way to estimate the size of the BLR by modelling the shape of the 
H${\alpha}$ emission line.  The broad and bright H${\alpha}$ emission line has been observed with unprecedented spectral and spatial resolution using the Space Telescope Imaging Spectrograph (STIS) aboard {\it HST} and modelling the line shape yielded an outer radius for the BLR of ${\sim}$ 1 pc \citep{Devereux2007} independent of whether the BLR gas is distributed in an inclined geometrically thin Keplerian disc or moving radially at the escape velocity. 
Consequently, M81 occupies a very unusual location in the BLR size-luminosity diagram 
by virtue of having a disproportionately large BLR for an such an extremely low luminosity active nucleus  \citep{Devereux2015}.

Such a large size for the BLR could be possible if the gas producing the broad Balmer lines is actually of much lower density than previously believed permitting the central UV--X-ray source to photoionize a single, large H${^+}$ region, rather like that observed around massive stars, planetary nebulae and supernova remnants. Motivation is provided by \cite{Heckman1980} who noted that the [O\,{\sc iii}] line emission is seemingly independent of both the [O\,{\sc i}] and [O\,{\sc ii}] in Liners and Seyferts as if the [O\,{\sc iii}] line emission arises as a separate entity, distinct from the narrow-line region.
A low density H${^+}$ region would, of course, be associated with a much higher ionization parameter than envisaged for high density broad-line clouds. Such an H${^+}$ region may even reconcile the emission lines seen in X-rays \citep{Page2003,Young2007} with the same ionized gas seen in UV and visible spectra \citep{Peimbert1981, Bruzual1982, Filippenko1988, Reichert1992, Ho1996}. 

Ambiguity surrounding the impact of reddening on the broad Balmer emission lines in M81 and the additional uncertainty associated with interpolating the unobservable H ionizing continuum between the UV and X-rays have led to sizable discrepancies between the number of ionizing photons expected to be produced by the central UV--X-ray source and the number required to excite the broad Balmer emission lines \citep{Bruzual1982, Ho1996, Devereux2007}. However, the matter is worth revisiting because \cite{Nemmen2014} have computed the intrinsic ionizing continuum expected for the advection dominated accretion flow (ADAF) that is believed to be at the heart of the Seyfert 1 nucleus in M81. Furthermore, the emergent spectrum
has been defined from the visible to the UV with unprecedented spatial resolution as a result of a recent spectroscopic program executed with  {\it HST}. Thus, it is timely to re-examine the enigma that M81 presents.  To this end, the new Space Telescope Imaging Spectrograph (STIS) observations are described in Section 2. A comparison of emission line measurements with the expectations  of various photoionization models are described in Section 3 followed by a discussion and conclusions in Sections 4 and 5, respectively.

\section{Observations}

Spectroscopic observations of the nucleus of M81 were acquired with STIS aboard {\it HST}  at the end of December 2017 during Cycle 25. Spectra were obtained using the 52 arcsec ${\times}$ 0.2 arcsec slit oriented at a position angle of 32{\degr} that is aligned neither with the radio-jet axis \citep{Bietenholz1996, Bietenholz2000} nor the kinematic line of nodes \citep{Devereux2003}. M81 was centred on the aperture using an on-board target acquisition peakup in the F28X50LP visible long-pass filter. The present manuscript will address the subset of PID 15123 observations, summarized in Table 1, that span the visible to the UV at the highest spectral resolution.

Several exposures were obtained with each grating. Consecutive exposures were dithered-along-the-slit in multiples of 0.1\arcsec.  Since the Space Telescope Science Data Analysis System (STSDAS) routine {\bf sshift} used to align dithered spectral images employs only integer pixel shifts, the shifted images are slightly misaligned by about one tenth of a pixel. The shifted CCD images were median combined, separately for each grating, using the STSDAS routine {\bf ocrreject} which effectively removed cosmic rays. Similarly, the aligned MAMA images were median combined, separately for each grating, using the STSDAS routine {\bf mscombine}.
The G140L spectra are unique in that they were obtained using two different integration times. Those images were weighted by exposure time yielding an average 952 s per exposure. Calibrated spectra were extracted from the median combined images, separately for each grating, using the  STSDAS routine {\bf x1d} employing a 7 pixel extraction width which includes ${\ge}$ 80\% of the encircled energy for an unresolved point source \citep{Proffitt2010}. Since the width of the slit is 0.2 arcsec the spectra presented in Figure 1 represent an integration over a 0.2 arcsec ${\times}$ 0.35 arcsec box for G750M and G430L with a smaller 0.2 arcsec ${\times}$ 0.175 arcsec box for G230L and G140L. The bright nucleus of M81 is unresolved in the visible and UV with an upper limit on the physical diameter of 0.7 pc for the continuum source \citep{Devereux1997a}.

\section{Results}

\begin{figure}
	\includegraphics[width=\columnwidth]{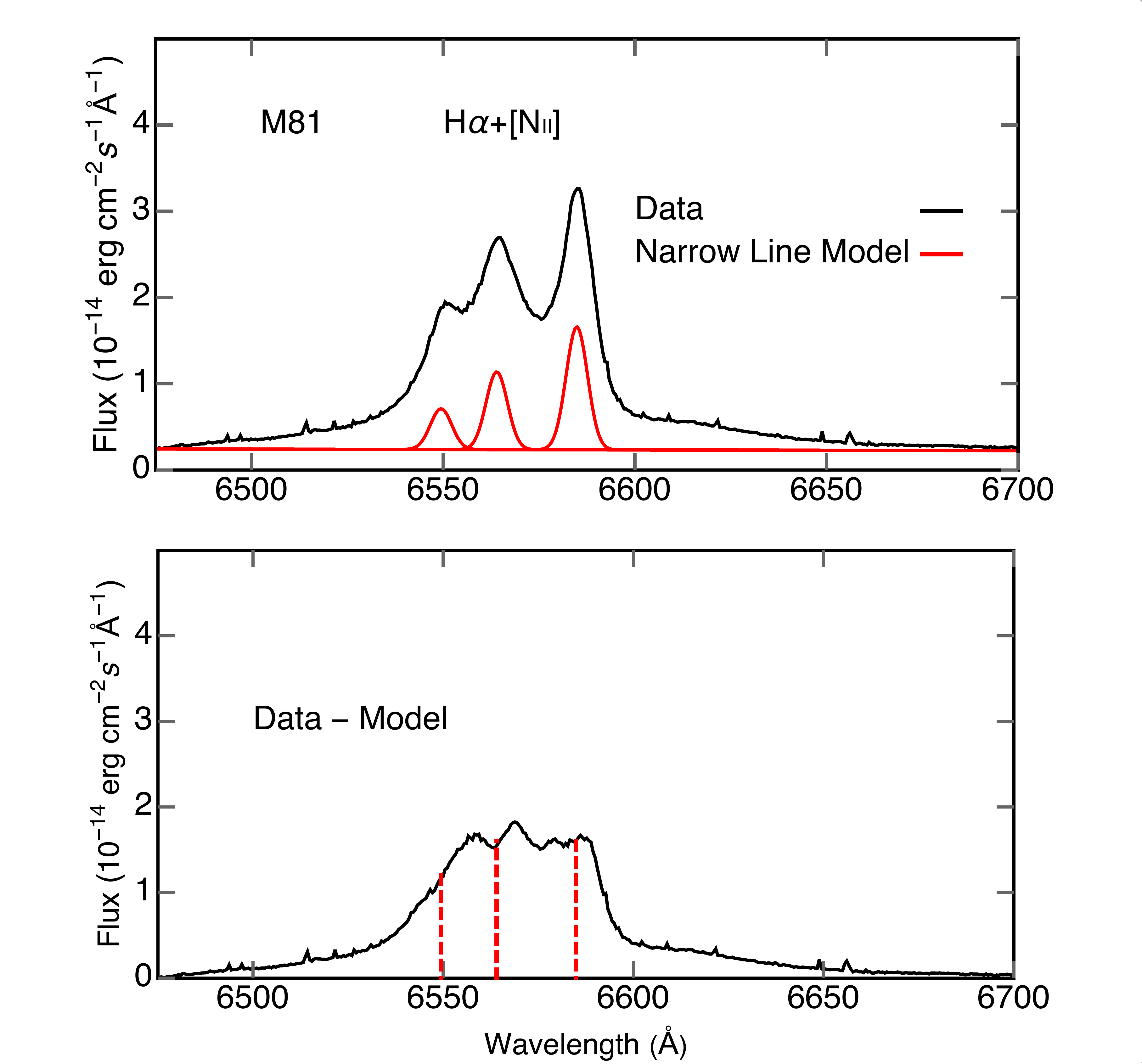}
    \caption{ Broad H${\alpha}$ emission line in M81. {\sl Top panel}: The observed spectrum
is shown in black and a model for the narrow lines and the underlying continuum is shown in red. {\sl Lower panel}: The broad 
H${\alpha}$ emission line profile
after the narrow lines have been subtracted. Dashed vertical red lines identify the central wavelengths of the subtracted lines.  }
    \label{fig:example_figure}
\end{figure}

\subsection{Measurement of Emission Lines}

The central wavelength, flux and full width at half-maximum (FWHM) was measured for the brightest and the least blended of the emission lines, identified in Figure 1, using the STSDAS contributed task {\bf specfit}. Results are listed in Table 2 together with their one sigma measurement uncertainties. 
The broad H${\alpha}$, and superimposed narrower {[}\ion{N}{ii}{]} emission lines, are illustrated in Figure 2. That the {[}\ion{N}{ii}{]} vacuum wavelength 6549.85 {\AA} and 6585.28 {\AA} emission lines can be clearly seen facilitates their subtraction. Atomic physics sets the wavelength of the fainter {[}\ion{N}{ii}{]} line relative to the brighter one, and constrains the flux to be 1/3 that of the brighter one, and requires that they share the same width. A narrow component of the vacuum wavelength 6564.61 {\AA} H${\alpha}$ emission line can also be seen (Figure 2) and was modelled using the same profile shape as employed for the {[}\ion{N}{ii}{]} emission lines, and scaled so as to not over-subtract the broad component. The velocity measured for the narrow line, -30 ${\pm}$ 11 km/s, is computed using the non-relativistic Doppler equation employing the wavelength difference; observed minus vacuum.

The three emission lines H${\beta}$, ${[}\ion{O}{iii}{]}$${\lambda}$${\lambda}$5008, 4960 are well resolved as illustrated in Figure 3. Encouragingly, the 
reddening insensitive ratio ${[}\ion{O}{iii}{]}$${\lambda}$$5008/{[}\ion{O}{iii}{]}$${\lambda}$4960 = 2.96 ${\pm}$ 0.57 agrees with the theoretical 
value \citep{Dimitrijevic2007}, albeit with a large measurement uncertainty. Although the narrow component of the H${\beta}$ line can not be uniquely
identified in Figure 3, one is presumably present. Consequently, a model was employed using the same shape and velocity as the narrow H${\alpha}$ line, 
and scaled so as to not over-subtract the broad H${\beta}$ line. The same procedure was 
adopted to obtain the flux associated with the narrow component of the H${\gamma}$ line. Ratios involving the narrow components of the Balmer lines determined 
this way are very similar to the corresponding ratios of the respective broad-line components.  An alternative scheme whereby the narrow components of H${\beta}$ 
and H${\gamma}$ are scaled relative to the narrow H${\alpha}$ line in proportion to the ratio expected for a dust-free photoionized gas \citep{Osterbrock1989} results 
in a very obvious over-subtraction of their respective broad lines, and was therefore not adopted.  

\begin{table*}
\centering
\caption{Emission Line Parameters }
\begin{threeparttable}
\begin{tabular}{|p{2cm}|p{2cm}|p{3cm}|p{2cm}|p{2cm}|}
\hline
Ion & Wavelength\tnote{\it a,b} &   Flux\tnote{\it b}   & FWHM\tnote{\it b} & V$_{cen}$\tnote{\it b} \\
& \AA &  10$^{-14}$ erg cm$^{-2}$ s$^{-1}$ & km s$^{-1}$ & km s$^{-1}$ \\
\hline
{[}\ion{S}{ii}{]}  & 6732.9 ${\pm}$ 0.1 & 6.6 ${\pm}$ 0.4 & 369 ${\pm}$ 26 & 10 ${\pm}$ 2 \\
{[}\ion{S}{ii}{]}  & 6718.5 ${\pm}$ 0.2 & 4.0 ${\pm}$ 0.1 & 354 ${\pm}$ 29  & 9 ${\pm}$ 8 \\
{[}\ion{N}{ii}{]}  & 6584.9 ${\pm}$ 0.3 & 10.0 ${\pm}$ 0.3  & 300 ${\pm}$ 20  & 20 ${\pm}$ 14 \\
H${\alpha}$ (broad) & 6571 ${\pm}$ 6 & 108 ${\pm}$ 3  & 2185 ${\pm}$ 25 & 192  \\
H${\alpha}$ (narrow) & 6564.0 ${\pm}$ 0.2 & 6.3 ${\pm}$ 0.2 & 300 & -30 ${\pm}$ 11 \\
{[}\ion{N}{ii}{]}  & 6549.4 ${\pm}$ 0.7 & 3.3 ${\pm}$ 1.0 & 300 & 33 \\
{[}\ion{O}{i}{]} & 6365.5 ${\pm}$ 0.1 & 5.3 ${\pm}$  0.2 & 619 ${\pm}$ 10 & -2 ${\pm}$ 4 \\
{[}\ion{O}{iii}{]}\tnote{\it c}  & 5011.1  ${\pm}$ 0.4  & 19.3  ${\pm}$  2.3  & 530 ${\pm}$ 11 & 174 ${\pm}$ 24 \\
{[}\ion{O}{iii}{]}\tnote{\it c} & 4961 ${\pm}$ 1 & 6.5 ${\pm}$ 1 & 596 ${\pm}$ 49 & 66 ${\pm}$ 21 \\
H${\beta}$ (broad) & 4866 ${\pm}$ 3 & 19.1 ${\pm}$ 0.5 & 1350 ${\pm}$ 140 & 208  \\
H${\beta}$ (narrow) & 4863 & 0.8  & 300 & -30  \\
{[}\ion{O}{iii}{]} & 4366 ${\pm}$ 3 &  2.8 ${\pm}$ 1.1  & 1261 ${\pm}$ 763 & 115 ${\pm}$ 371  \\
H${\gamma}$ (broad) & 4345 ${\pm}$ 4 & 5.8 ${\pm}$ 1.1 & 1328 ${\pm}$ 140 & 208  \\
H${\gamma}$ (narrow) & 4342  & 0.3 & 300 & -30 \\
H${\delta}$ (broad) & 4103 ${\pm}$ 1 & 3.5 ${\pm}$ 0.8 & 1100 ${\pm}$ 140 & 219 \\
{[}\ion{S}{ii}{]}\tnote{\it d} & 4073  ${\pm}$ 0.8 & 6.6  ${\pm}$ 0.3 & 962 ${\pm}$ 128 & -39 ${\pm}$ 57 \\
{[}\ion{O}{ii}{]}\tnote{\it d} & 3731 ${\pm}$ 0.3 &  5.3 ${\pm}$ 0.3 & 760 ${\pm}$ 16 & 194 ${\pm}$ 24 \\
\ion{Mg}{ii}\tnote{\it e}  & 2798.0  ${\pm}$ 0.5  &  56 ${\pm}$ 1 & 2292 ${\pm}$ 107 & 0 ${\pm}$ 50 \\
Fe? & 2621 & 3.6 ${\pm}$ 0.1 & ... & ... \\
{[}\ion{O}{ii}{]}  & 2471.5 ${\pm}$ 0.1 & 2.9 ${\pm}$ 0.1 & 1191 ${\pm}$ 41 & 66 ${\pm}$ 16\\
\ion{C}{ii}{]}\tnote{\it f} & 2328.0 ${\pm}$ 0.3 & 20 ${\pm}$ 1 & 1960 ${\pm}$ 85 & -31 ${\pm}$ 34 \\
\ion{N}{ii}{]} & 2142.8 ${\pm}$ 0.2 & 1.8 ${\pm}$ 0.05 & 1695 ${\pm}$ 79 & -91 ${\pm}$ 34 \\
\ion{C}{iii}{]}\tnote{\it c} & 1908.0 ${\pm}$ 0.4 & 30 ${\pm}$ 1 & 1995 ${\pm}$ 162 & -115 ${\pm}$ 65 \\
{[}\ion{Ne}{iii}{]} & 1814 ${\pm}$ 0.6 & 2.7 ${\pm}$ 0.2 & 3393 ${\pm}$ 223 & -112 ${\pm}$ 100 \\
\ion{N}{iii}{]} & 1751.9 ${\pm}$ 0.4 & 4.2 ${\pm}$ 0.2 & 2435 ${\pm}$ 185  & 70 ${\pm}$ 67 \\
\ion{O}{iii}{]} & 1664.1 ${\pm}$ 0.4 & 2.4 ${\pm}$ 0.3 & 1500 ${\pm}$ 139 & -9 ${\pm}$ 69 \\
\ion{He}{ii}\tnote{\it g}  & 1642  ${\pm}$ 2  &  6.4 ${\pm}$ 0.2  & 4296 ${\pm}$ 518  & 310 ${\pm}$ 405 \\
\ion{C}{iv}\tnote{\it e} & 1549.4 ${\pm}$ 0.3 & 24 ${\pm}$ 1 & 4148 ${\pm}$ 135 & 231 ${\pm}$ 52 \\
\ion{N}{iv}{]} & 1486 ${\pm}$ 0.25 & 2.9 ${\pm}$ 0.1 & 2529 ${\pm}$ 160 & -101 ${\pm}$ 48 \\
\ion{Si}{iv}+\ion{O}{iv}{]} & 1400 ${\pm}$ 0.15 & 7.0 ${\pm}$ 0.2 & 4404 ${\pm}$ 74  & 95 ${\pm}$ 32 \\

\hline

\end{tabular}

\begin{tablenotes}\footnotesize

\item[\it a] observed central wavelength
\item[\it b] values without uncertainties are fixed quantities.
\item[\it c] sum of a logarithmic and a gaussian line profile with the flux apportioned 10:1, respectively. The model created to describe the 
{[}\ion{O}{iii}{]} ${\lambda}$4364 emission line is noticeably wider and has a different shape than {[}\ion{O}{iii}{]} ${\lambda}$5008. The former is best described 
by a gaussian whereas the latter is better represented by the sum of logarithmic and gaussian components of different FWHM and central wavelength. 
\item[\it d] unresolved doublets
\item[\it e] uncertain due to self-absorption.
\item[\it f] uncertain continuum level.
\item[\it g] sum of two gaussians of approximately the same brightness, but different FWHM and central wavelength. 
\end{tablenotes}
\end{threeparttable}
\end{table*}

\begin{figure}
	\includegraphics[width=\columnwidth]{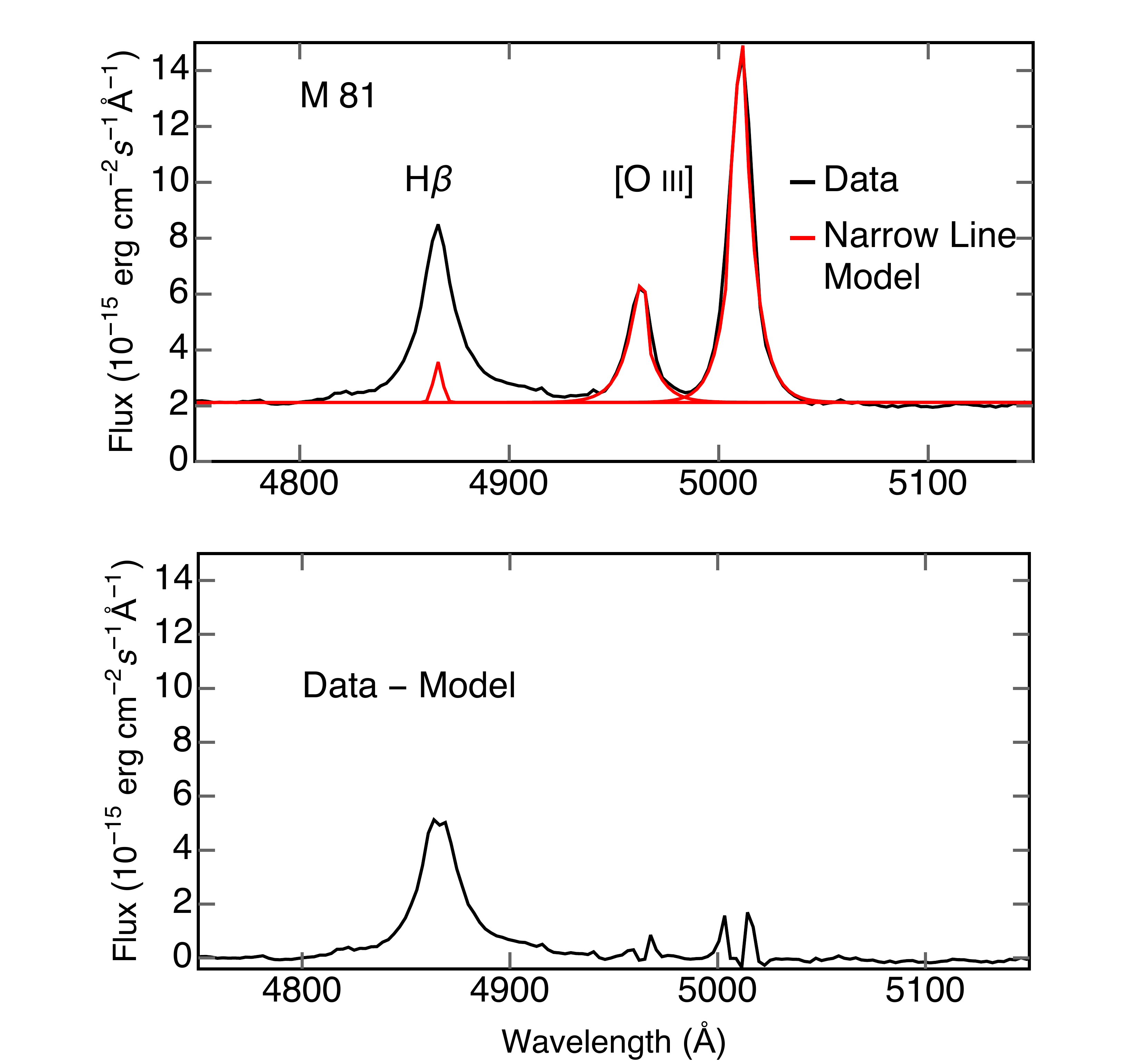}
    \caption{ Broad H${\beta}$ emission line in M81. {\sl Top panel}: The observed spectrum
is shown in black and a model for the narrow line H${\beta}$, the {[}\ion{O}{iii}{]} lines, and the underlying continuum is shown in red. {\sl Lower panel}: The broad 
H${\beta}$ emission line profile after the model lines have been subtracted. }
    \label{fig:example_figure}
\end{figure}

\begin{figure}
	\includegraphics[width=\columnwidth]{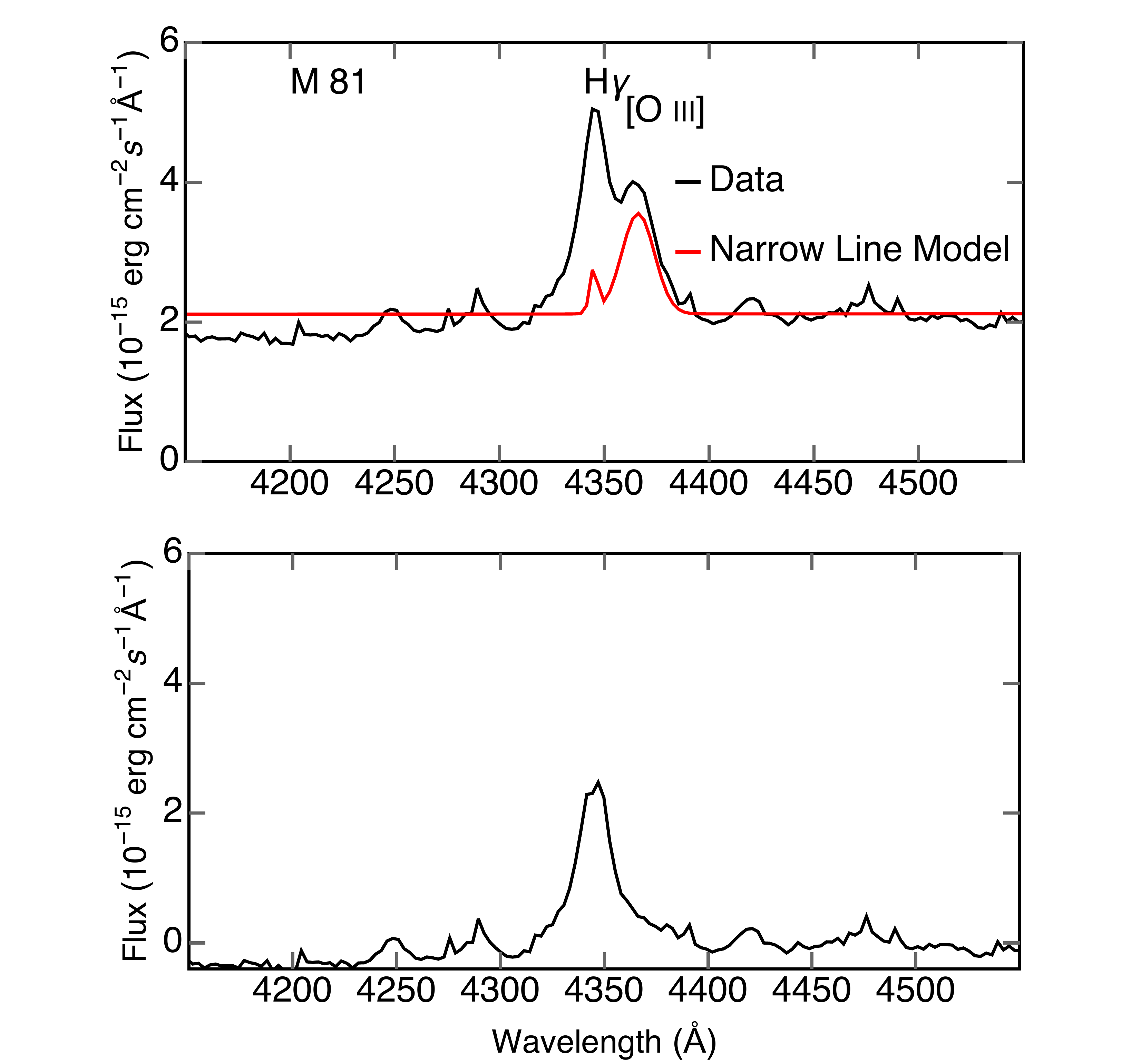}
    \caption{ Broad H${\gamma}$ emission line in M81. {\sl Top panel}: The observed spectrum
is shown in black and a model for the narrow H${\gamma}$ line, the {[}\ion{O}{iii}{]} line and underlying continuum is shown in red. {\sl Lower panel}: The broad H${\beta}$ emission line profile after the model lines have been subtracted. }
    \label{fig:example_figure}
\end{figure}

\begin{figure}
	\includegraphics[width=\columnwidth]{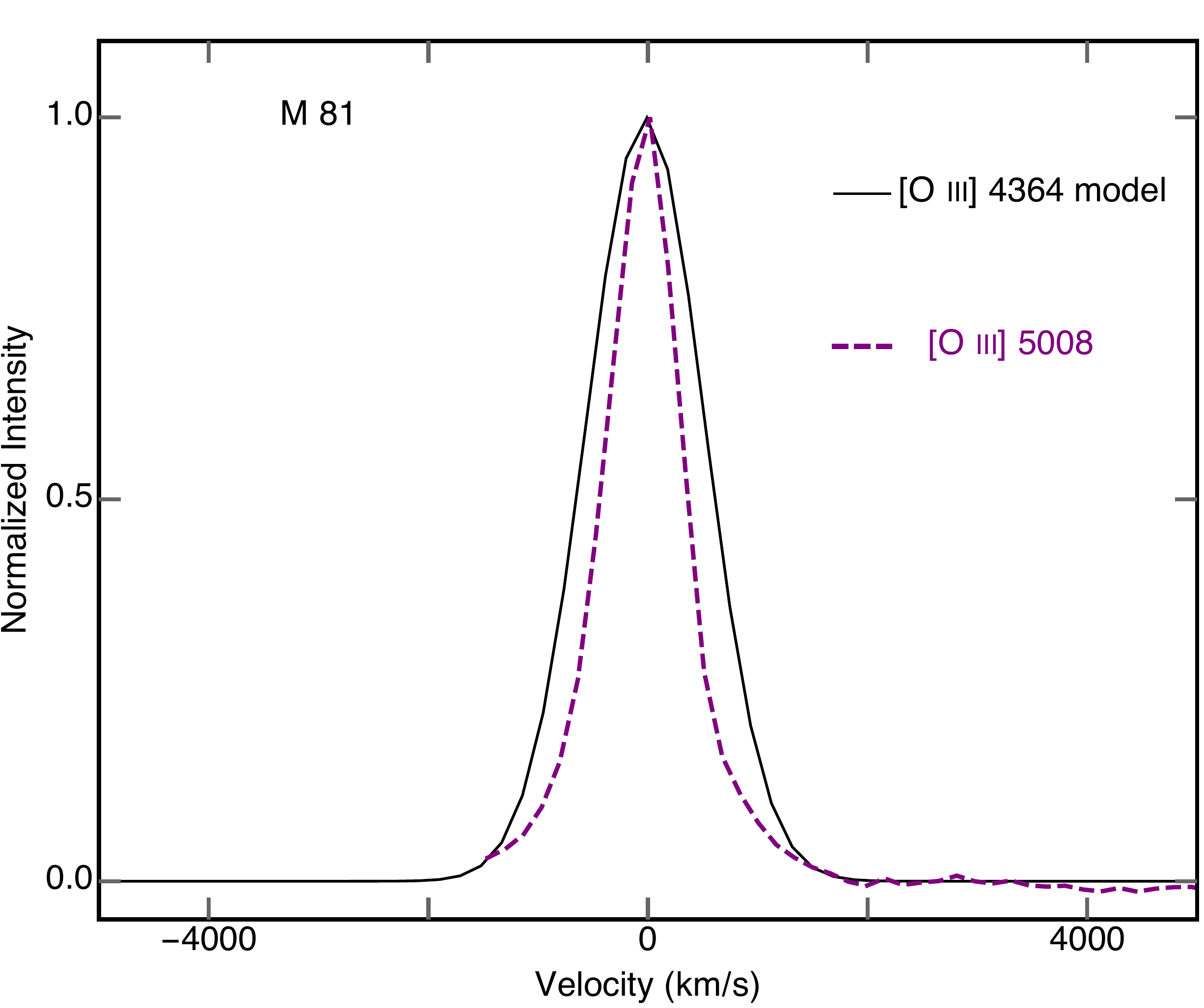}
    \caption{ Comparison of the observed ${[}\ion{O}{iii}{]}$${\lambda}$5008 emission line (purple, dashed) with the model employed for ${[}\ion{O}{iii}{]}$${\lambda}$4364 (black, solid). The latter appears slightly wider in rest-frame velocity space.}
    \label{fig:example_figure}
\end{figure}

Figure 4 shows that broad H${\gamma}$ is severely blended with what appears to be a broadened vacuum wavelength 4364.43 {\AA}
${[}\ion{O}{iii}{]}$ emission line. The model that was required to cleanly subtract ${[}\ion{O}{iii}{]}$${\lambda}$4364 from H${\gamma}$ is slightly wider than
the observed ${[}\ion{O}{iii}{]}$${\lambda}$5008 profile, as illustrated in Figure 5, although the difference in FWHM is
barely significant, 530 ${\pm}$ 11 km s$^{-1}$ for [O\,{\sc iii}]${\lambda}$5008 compared to 1261 ${\pm}$ 763 km s$^{-1}$ for [O\,{\sc iii}]${\lambda}$4364,
consistent with prior measurements \citep{Filippenko1988, Ho1996}. The wider ${[}\ion{O}{iii}{]}$${\lambda}$4364 line is plausible given its higher critical  
density, 3.1 ${\times}$ 10$^7$ cm$^{-3}$, compared to
8.5 ${\times}$ 10$^5$ cm$^{-3}$ for [O\,{\sc iii}]${\lambda}$5008. Critical densities were computed for all the forbidden lines listed in Table 2 using the STSDAS
task {\bf ionic} adopting an electron temperature of 2 ${\times}$ 10$^4$ K. A significant correlation, reported previously by \citet{Filippenko1988}, between FWHM
and critical density for forbidden and semi-forbidden lines, is illustrated using the new data in Figure 6, supporting their explanation that the marginal difference between the FWHM
of the [O\,{\sc iii}] lines is caused by a density gradient.
Interestingly, there are many broad forbidden emission lines in the UV, and including them extends the correlation by about three orders of magnitude in critical density. 
In principle, the ratio ${[}\ion{O}{iii}{]}$(${\lambda}$5008 + ${\lambda}$4960)/${\lambda}$4364, provides a measure of electron temperature in the O$^{++}$ region. 
More conveniently expressed as 4/3 ${[}\ion{O}{iii}{]}$(${\lambda}$5008/${\lambda}$4364), the ratio has an observed value of 9 ${\pm}$ 3, consistent with
prior measurements made with ground based telescopes \citep{Peimbert1981, Filippenko1988, Ho1996}. The extremely low 
${[}\ion{O}{iii}{]}$${\lambda}$5008/${\lambda}$4364 ratio measured for the nucleus of M81 implicates collisionally excited ions radiating at or near the critical
density for those transitions \citep{Filippenko1988}, or photoionized gas of low metallicity and high ionization parameter  \citep[][and references therein]{Binette1996}.
Alternatively, gas photoionized by fast shocks \citep{Dopita1995,Morse1996}. 

\begin{figure}
	\includegraphics[width=\columnwidth]{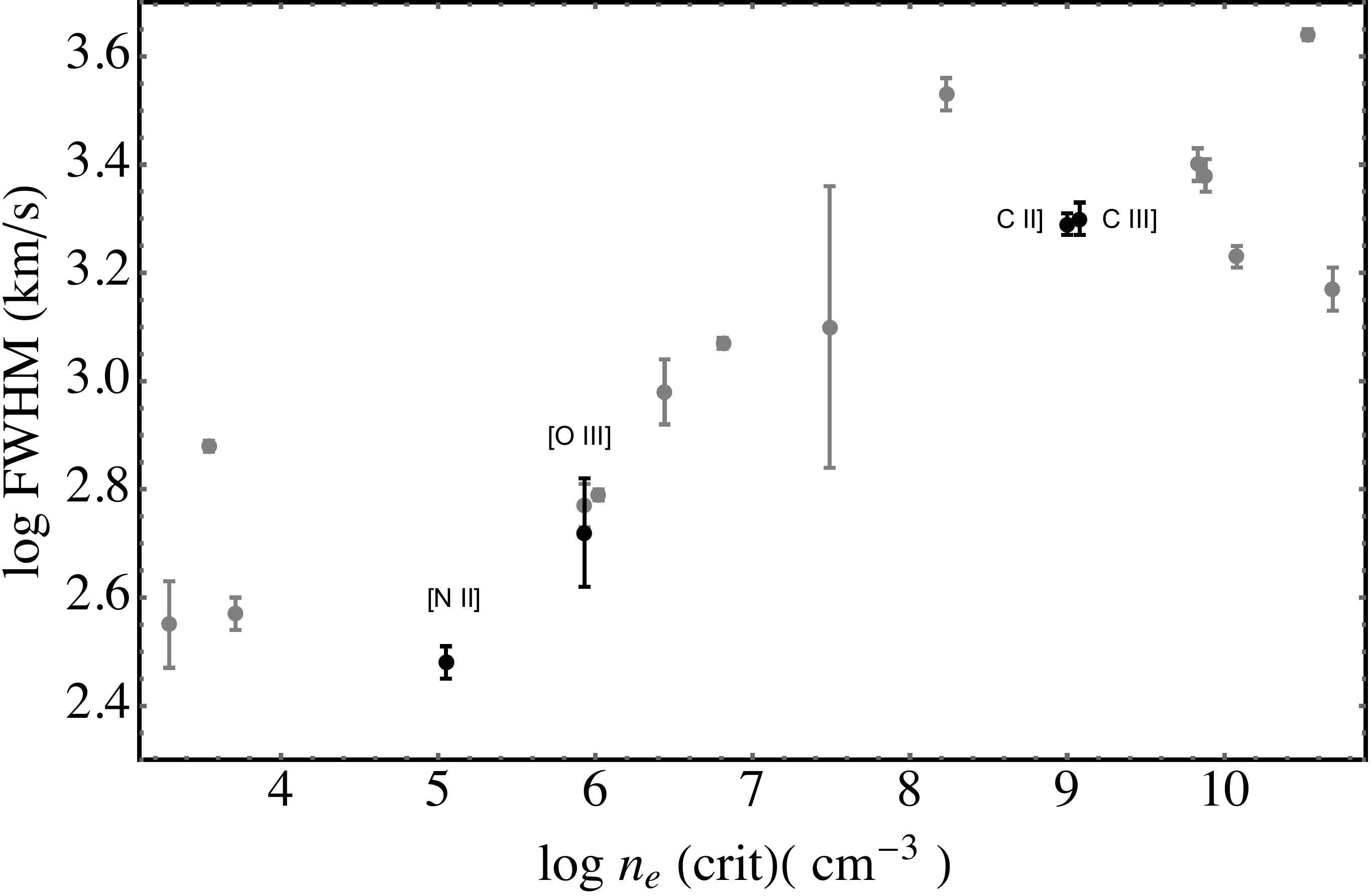}
    \caption{The ordinate depicts the FWHM of the emission line in km s$^{-1}$, whereas the abscissa depicts the critical electron density of the transition in units of electrons/cm$^{3}$. Labelled black dots distinguish bright forbidden and semi-forbidden emission lines, ${\ge}$ 10$^{-13}$ erg cm$^{-2}$ s$^{-1}$, from fainter ones.   }
    \label{fig:example_figure}
\end{figure}

\begin{figure}
	% To include a figure from. a file named example.*
	\includegraphics[width=\columnwidth]{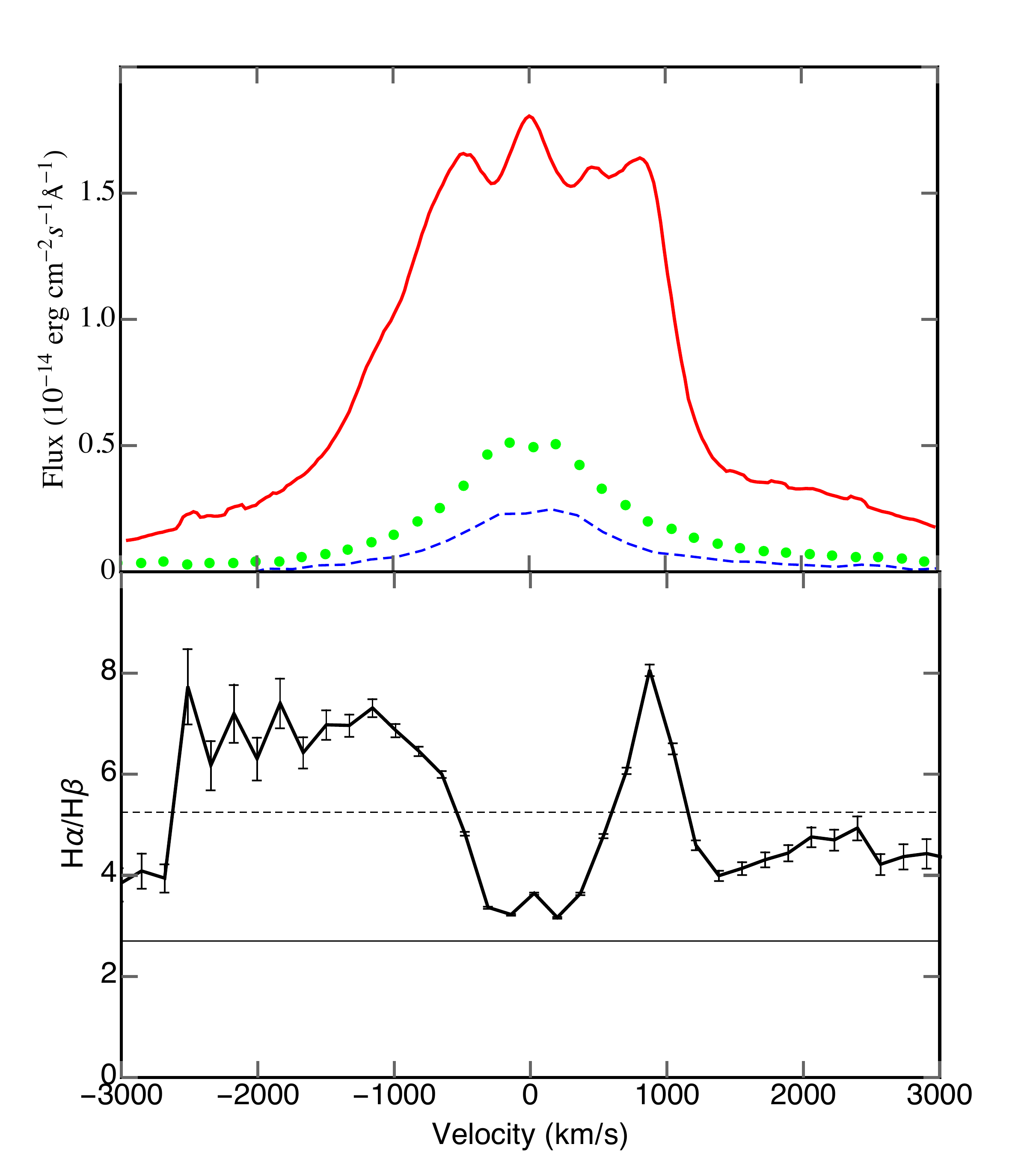}
    \caption{(Top panel) Emission line profiles for H${\alpha}$ (red-solid line) smoothed to the resolution of H${\beta}$ (green-dotted line)
    and H${\gamma}$ (blue-dashed line). The ordinate indicates observed flux in units of erg cm$^{-2}$ s$^{-1}$ {\AA}$^{-1}$.  (Lower panel) The ordinate indicates the observed H${\alpha}$/H${\beta}$ flux ratio. The solid horizontal black line represents the ratio 2.7 expected for photoionization.The horizontal dashed line identifies the observed unweighted mean ratio 5.3. Error bars denote flux measurement uncertainties only. The abscissa represents radial velocity in km s$^{-1}$ relative to the rest frame of the broad H${\alpha}$ line (see Section 3.3 for details). }

    \label{fig:example_figure}
\end{figure}

Figure 3 illustrates that ${[}\ion{O}{iii}{]}$${\lambda}$5008 is slightly brighter than broad H${\beta}$, but it is also narrower causing the ratio of the flux in the two lines to be 1 ${\pm}$ 0.1, also consistent with prior ground based measurements. The consistency arises because the majority 85 ${\pm}$ 13\% of the ${[}\ion{O}{iii}{]}$${\lambda}$5008 flux measured with terrestrial telescopes originates from the central 0.2 arcsec ${\times}$ 0.35 arcsec region measured with STIS. 
The reddening insensitive ${[}\ion{O}{iii}{]}$${\lambda}$5008/H${\beta}$ ratio has been interpreted variously.  Photoionization models show that it depends on both metallicity and ionization parameter \citep{Ferland1983, Binette1985}. ${[}\ion{O}{iii}{]}$${\lambda}$5008/H${\beta}$ has been used in conjunction with ${[}\ion{O}{iii}{]}$${\lambda}$5008/${\lambda}$4364 to distinguish shocks from their associated H${^+}$ regions \citep{Dopita1995, Dopita1996}.
${[}\ion{O}{iii}{]}$${\lambda}$5008/H${\beta}$ has been used in conjunction with other line ratios to segregate various types of AGN \citep[e.g.][and references therein]{Baldwin1981, Veilleux1987, Kewley2006, Feltre2016}.  ${[}\ion{O}{iii}{]}$${\lambda}$5008/H${\beta}$ and ratios of other lines qualifies the active nucleus in M81 to be a Liner according to the diagnostic diagrams of \citet{Kewley2006} and a Seyfert 1.2 according to the scheme of \cite{Whittle1992}. 
The duplicity surrounding the true nature of the AGN in M81 warrants further investigation.

\subsection{Evidence for Shocked Gas}
The UV spectra illustrated in Figure 1 reveal a plethora of forbidden and semi-forbidden emission lines, several of which are rarely seen including 
\ion{Si}{iv}+\ion{O}{iv}{]} ${\lambda}$1400, \ion{O}{iii}{]} ${\lambda}$1664, \ion{N}{iii}{]} ${\lambda}$1751, \ion{Ne}{iii}{]} ${\lambda}$1814, \ion{N}{ii}{]} ${\lambda}$2143, 
and {[}\ion{O}{ii}{]} ${\lambda}$2470. Additionally, bright but absorbed Ly${\alpha}$ ${\lambda}$1215, \ion{C}{iv} ${\lambda}$1549, \ion{C}{iii}{]} ${\lambda}$1908, 
\ion{C}{ii}{]} ${\lambda}$2328, and \ion{Mg}{ii} ${\lambda}$2798. Many of these same lines were predicted to be bright in shocked ionized gas by
\citet{Shull1979} but it was \cite{Dopita1995,Dopita1996} who illustrated the utility of the {[}\ion{O}{iii}{]}${\lambda}$5008/${\lambda}$4364 versus 
{[}\ion{O}{iii}{]}${\lambda}$5008/H${\beta}$ diagram in segregating pure shocks from those associated with bright precursor H${^+}$ regions. 
The observed {[}\ion{O}{iii}{]}${\lambda}$5008/${\lambda}$4364 and {[}\ion{O}{iii}{]} ${\lambda}$5008/H${\beta}$ ratios locate M81 unambiguously in 
the ``shock only" region of their diagram \citep[][see their Figure 7]{Dopita1995}. The evidence for shocks is discussed further in Section 4.

\subsection{A Redshifted BLR?}

\citet{Peimbert1981} reported that the broad H${\alpha}$ emission line is redshifted with respect to the narrow component.  
The broad H${\alpha}$ emission line is essentially rectangular in shape with a wide and flat top that limits the accuracy with which 
the wavelength of the line centre can be determined (Table 2). However,  a residual peak at what appears to be the bisector of the broad 
H${\alpha}$ emission line is redshifted by 4.2 {\AA} relative to the subtracted narrow component (see Figure 2), in good agreement with a 
prior report by \citet{Peimbert1981}. The redshift associated with the 
broad H${\beta}$ and H${\gamma}$ lines is more difficult to discern in the lower resolution G430L spectra, but imposing a redshift of 3 {\AA} on those lines 
with respect to their vacuum wavelengths resulted in their alignment with each other and H${\alpha}$ (Figure 7). Interestingly, the average doppler shift 
measured with respect to vacuum wavelength collectively for the {[}\ion{N}{ii}{]} and narrow H${\alpha}$, plus the {[}\ion{S}{ii}{]} and {[}\ion{O}{i}{]} lines flanking 
the broad H${\alpha}$ line, is 0.0 ${\pm}$ 0.4 {\AA}. This is to be compared with prior observations of the more extended narrow-line region (NLR) which indicated 
a 1 {\AA} blueshift based on the brighter {[}\ion{N}{ii}{]} line \citep{Devereux2003}. 
A more complete picture of the emission line kinematics is presented in Figure 8 where the velocity of the line centre, 
V$_{cen}$, expressed in km s$^{-1}$ and computed using the non-relativistic Doppler equation employing the wavelength difference; observed minus vacuum, is plotted against wavelength for 29 emission lines listed in Table 2 (excluding the one emission line with an uncertain identification at 2621 {\AA}). 
The figure shows that the emission lines exhibit a wide range in V$_{cen}$, from a blueshift of -58 km s$^{-1}$ to a redshift of 174 km s$^{-1}$, with no dependence 
on wavelength or line brightness. Emission lines with redshifts similar to the broad Balmer lines include ${[}\ion{O}{iii}{]}$ ${\lambda}$${\lambda}$5008, 4364, 
the ${[}\ion{O}{ii}{]}$ ${\lambda}$${\lambda}$3727, 3729 blend, \ion{He}{ii} ${\lambda}$1640 and \ion{C}{iv} ${\lambda}$1548, but they can hardly be described as outliers.
Collectively, the  mean V$_{cen}$ = 58 ${\pm}$ 116 km s$^{-1}$, the dispersion being approximately a factor of two larger than the ${\pm}$ 60 km s$^{-1}$ 
amplitude measured for the brighter {[}\ion{N}{ii}{]} emission line reported in \cite{Devereux2003}. By comparison, the line-centre velocities and associated dispersion 
are dwarfed by the much larger FWHM which spans an impressive range from 300 km s$^{-1}$ to 4500 km s$^{-1}$ that is correlated with wavelength in the sense 
that the broadest lines occur in the UV as illustrated in Figure 9. The broadest emission line is actually a blend of \ion{Si}{iv}+\ion{O}{iv}{]} at 1400 {\AA},
closely followed by \ion{He}{ii} ${\lambda}$1642 and \ion{C}{iv} ${\lambda}$1549. The new observations reveal no compelling evidence for a redshifted BLR
because there is no correlation between FWHM and V$_{cen}$ as illustrated in Figure 10. Furthermore, naively identifying the BLR with
{\it the region emitting the broadest emission lines} would inevitably include many broad forbidden lines as well. Thus, any model for the BLR in M81 must include 
an explanation for the broad forbidden lines as well as the broad permitted lines.

\begin{figure}
	% To include a figure from. a file named example.*
	\includegraphics[width=\columnwidth]{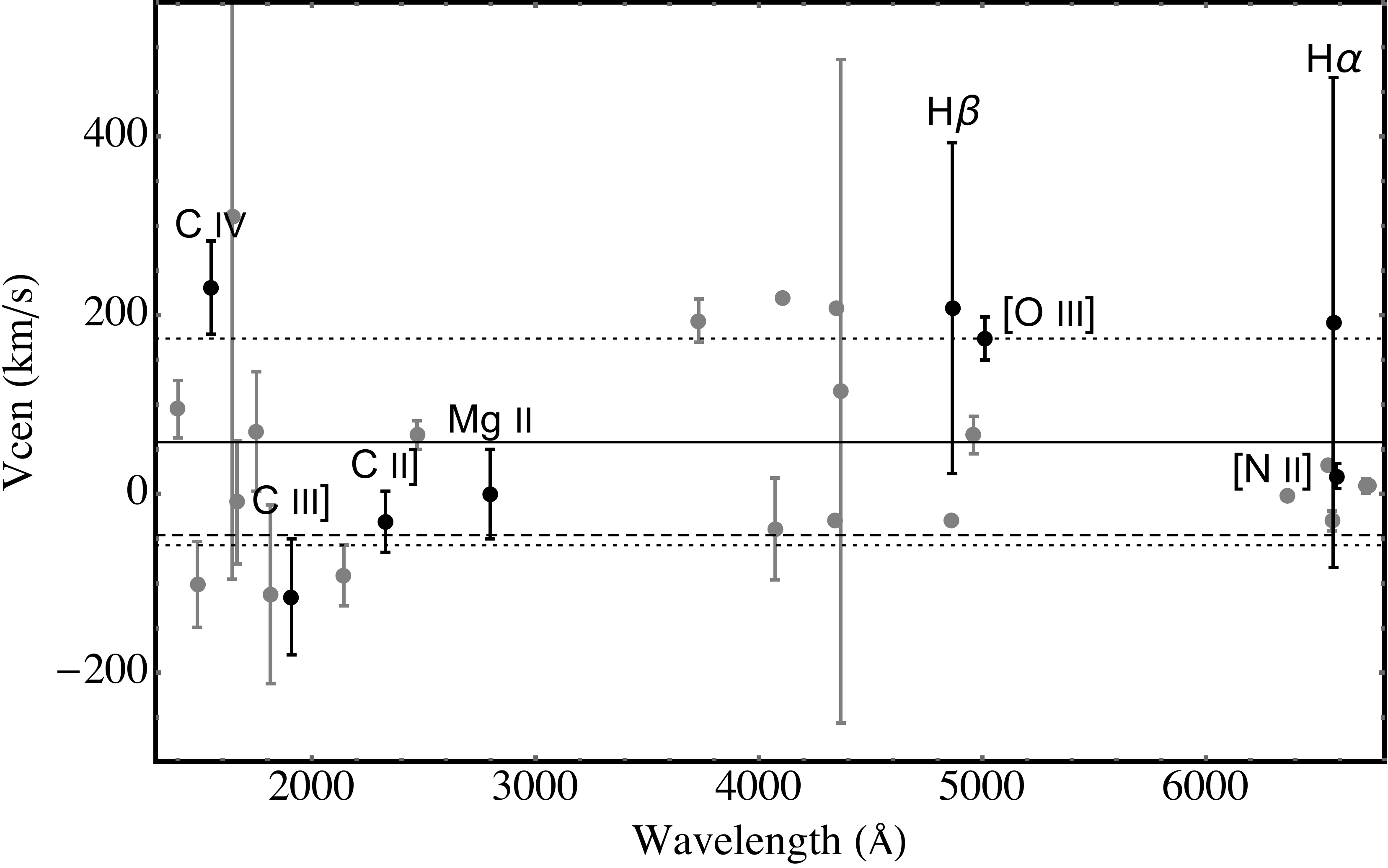}
    \caption{The ordinate depicts the velocity of the emission line centre in km s$^{-1}$, whereas the abscissa depicts wavelength in units of Angstroms. Labelled black dots distinguish bright emission lines, ${\ge}$ 10$^{-13}$ erg cm$^{-2}$ s$^{-1}$, from fainter ones.  The horizontal solid and dotted lines depict the mean and ${\pm}$ 1 ${\sigma}$ standard deviation for the distribution, respectively. The black dashed line represents the blueshift measured for the larger scale NLR (see Section 3.3 for details).  }

    \label{fig:example_figure}
\end{figure}

\begin{figure}
	\includegraphics[width=\columnwidth]{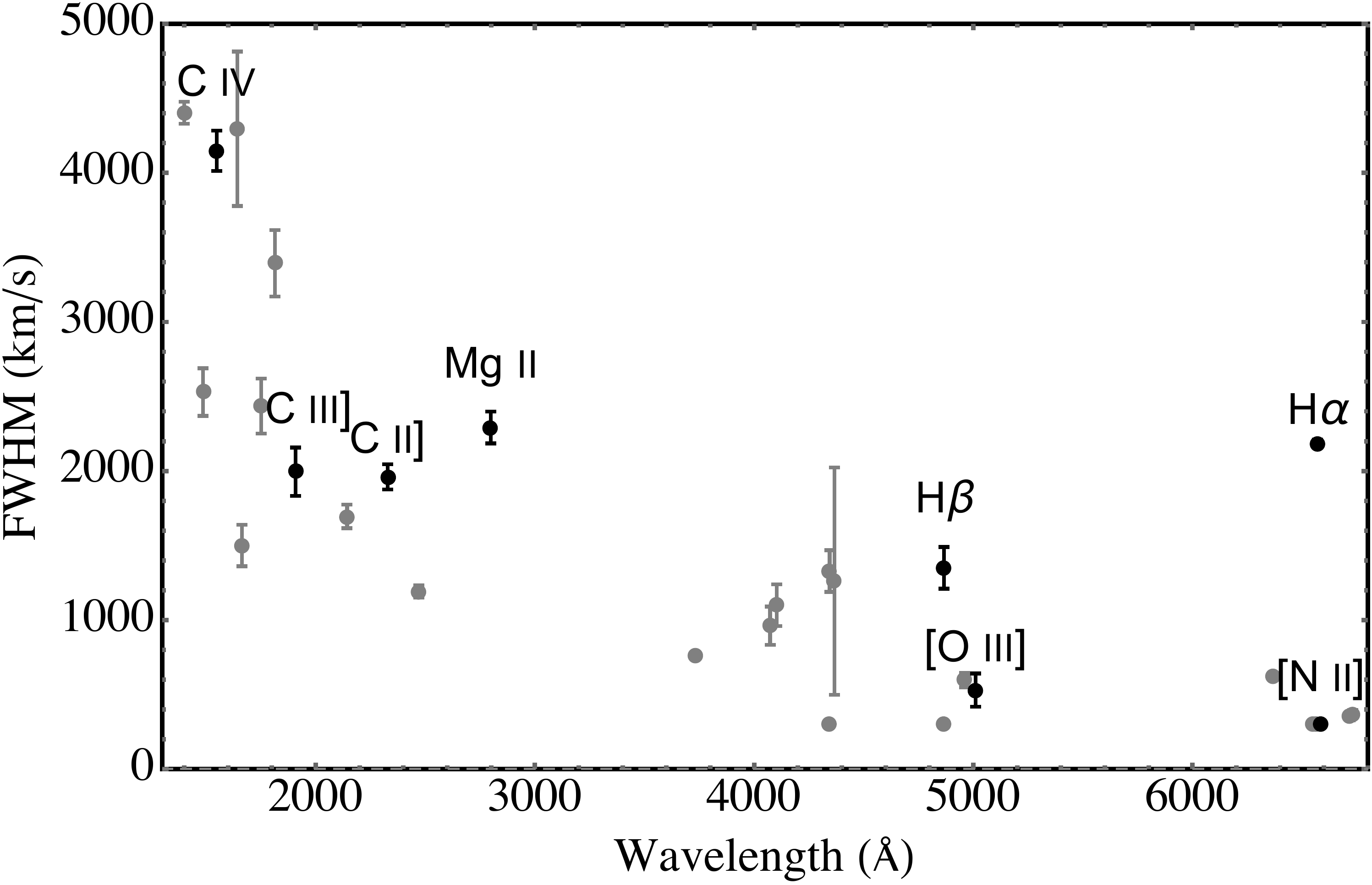}
    \caption{The ordinate depicts the FWHM of the emission line in km s$^{-1}$, whereas the abscissa depicts wavelength in units of Angstroms. Labelled black dots distinguish bright emission lines, ${\ge}$ 10$^{-13}$ erg cm$^{-2}$ s$^{-1}$, from fainter ones.   }
    \label{fig:example_figure}
\end{figure}

\begin{figure}
	\includegraphics[width=\columnwidth]{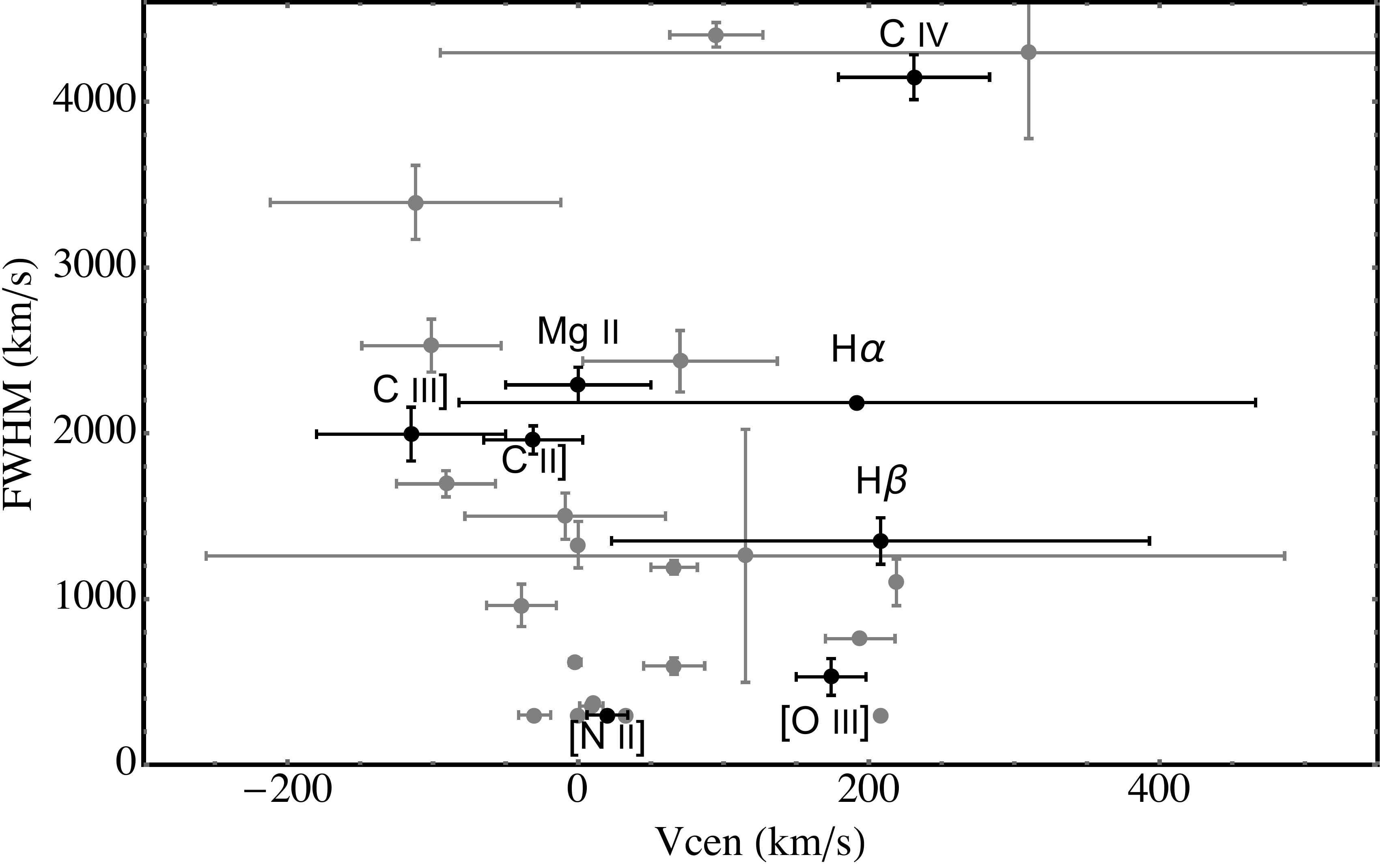}
    \caption{The ordinate depicts the FWHM of the emission line in km s$^{-1}$, whereas the abscissa depicts the velocity of the emission line centre in km s$^{-1}$. Labelled black dots distinguish bright emission lines, ${\ge}$ 10$^{-13}$ erg cm$^{-2}$ s$^{-1}$, from fainter ones.   }
    \label{fig:example_figure}
\end{figure}

\begin{figure}
	\includegraphics[width=\columnwidth]{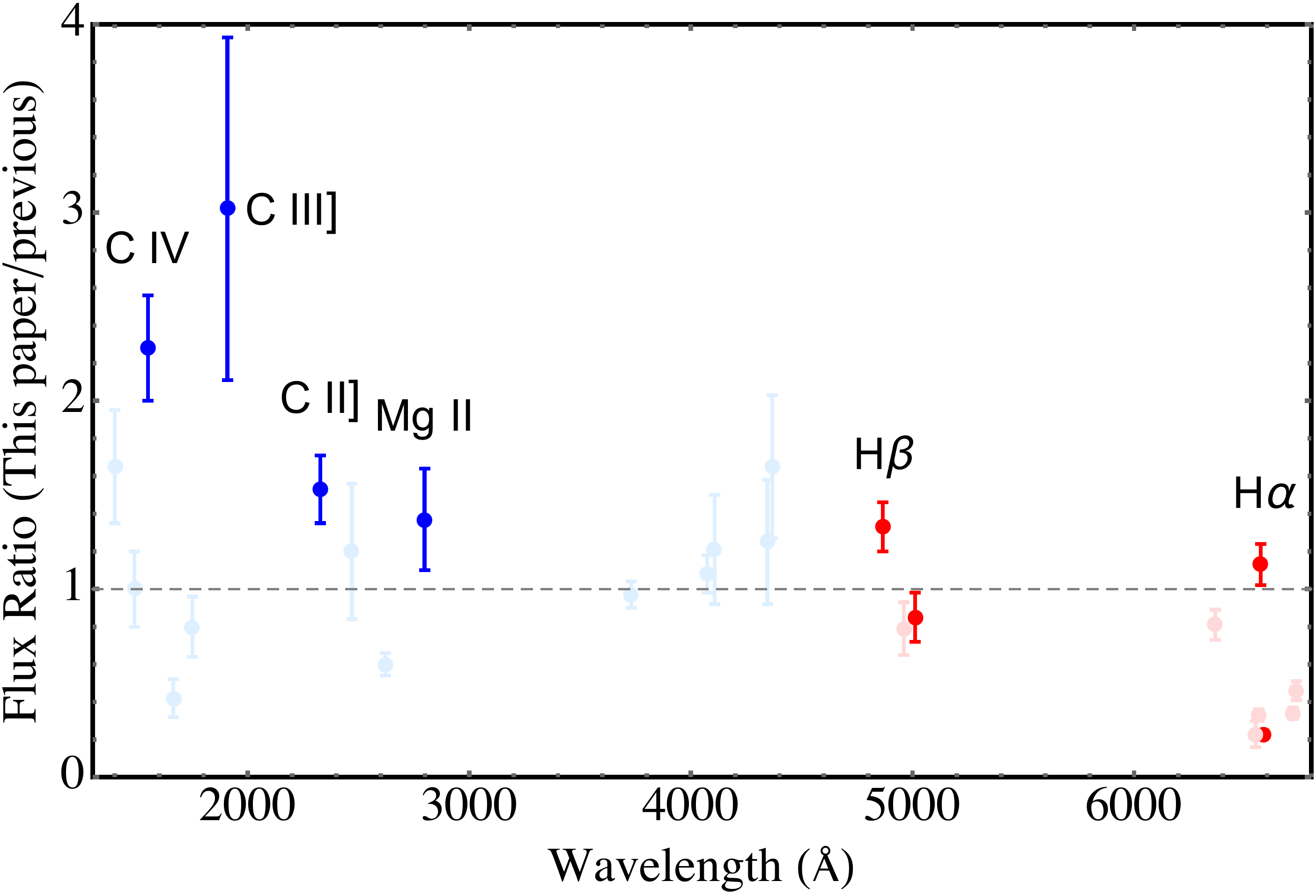}
    \caption{Comparison with previous measurements of emission line flux. The dimensionless ratio, plotted on the ordinate, represents the STIS measurement (this paper) divided by a prior measurement for emission lines in common with \citet{Ho1996}. The abscissa depicts wavelength in units of Angstroms. Red dots and blue dots identify ground based and space based measurements, respectively, occurring in the denominator of the ratio. Bright dots distinguish bright emission lines, ${\ge}$ 10$^{-13}$ erg cm$^{-2}$ s$^{-1}$, from fainter ones. Bright emission lines that may plausibly be brighter than previously observed are labelled.}
    \label{fig:example_figure}
\end{figure}

\subsection{Comparison with Previous Measurements}

A comprehensive list of emission line measurements for M81 was presented previously by \citet{Ho1996} who, using an independent set of observations, measured fluxes for all of the same lines reported in Table 2 with 
the exception of \ion{He}{ii} ${\lambda}$1640, \ion{N}{ii}{]} ${\lambda}$2143, and {[}\ion{Ne}{iii}{]} ${\lambda}$1814. An interesting comparison 
is illustrated in Figure 11, presented as the ratio of the flux reported in Table 2, divided by that of \citet{Ho1996}, and segregated by emission line wavelength. At the longest wavelengths there is a group of bright lines ({[}\ion{N}{ii}{]}, {[}\ion{S}{ii}{]} and narrow H${\alpha}$) for which the ratio is < 1, understandably so, because the STIS observations employed a narrower slit than the ground based observations of \citet{Ho1996} indicating that these narrow emission lines arise from a spatially extended region. More intriguing are the bright lines that are associated with ratios > 1 which includes the broad emission lines of H${\alpha}$, H${\beta}$, and vacuum wavelength \ion{Mg}{ii} ${\lambda}$2798, \ion{C}{ii}{]} ${\lambda}$2328, \ion{C}{iii}{]} ${\lambda}$1908 and \ion{C}{iv} ${\lambda}$1548. The trend is for the ratio to increase towards the UV. However, only \ion{C}{iv} has brightened significantly, by ${\sim}$ 230\%. A detailed comparison of the \ion{C}{iv} emission line with that reported by \citet{Ho1996} is presented in Figure 12 illustrating that the emission line appears to have retained much of its shape by having brightened over a wide range of velocities, but the adjacent continuum has not changed at all, and is the same, to within 10\%, as that measured in a prior {\it HST} image \citep{Devereux1997a}.

\begin{figure}
	\includegraphics[width=\columnwidth]{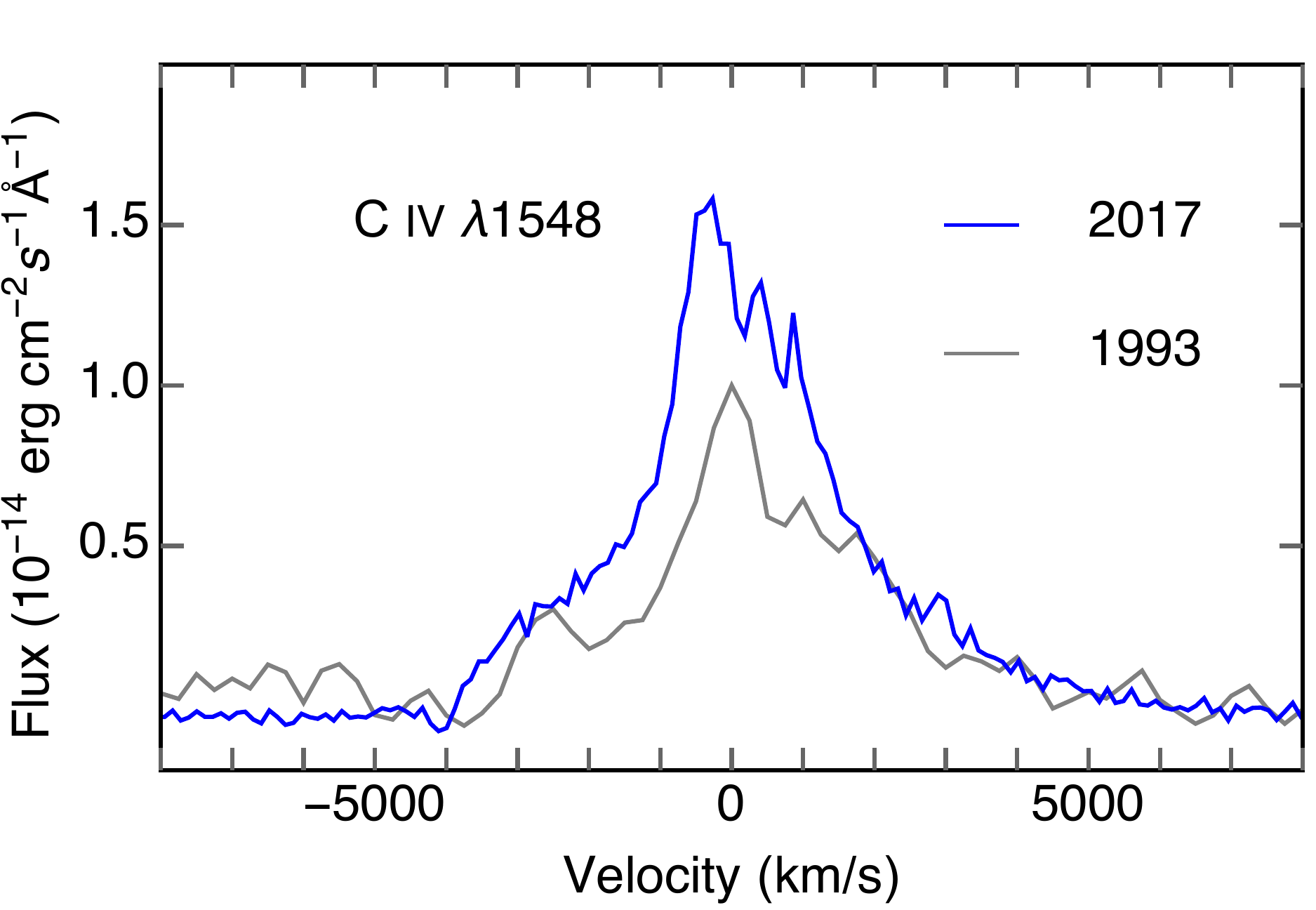}
    \caption{Comparison of the C\,{\sc iv} ${\lambda}$1548 emission line profiles measured with the Faint Object Spectrograph in 1993 with the more recent STIS observation in 2017. The ordinate measures flux in units of erg cm$^{-2}$ s$^{-1}$ {\AA}$^{-1}$ whereas the abscissa is in units of km s$^{-1}$ in the rest frame of M81.}
    \label{fig:example_figure}
\end{figure}

Independent measurements of the broad H${\alpha}$ emission line flux, based on {\it HST} observations, are in good agreement \citep{Bower1996, Devereux2003, Balmaverde2014}. Additionally,
\citet{Filippenko1988} noted that the broad H${\alpha}$ emission line flux has remained constant for the duration of ground based observations spanning several years.
The double-peaked profile reported by \citet{Bower1996} was evidently transient having not been seen in any subsequent spectra obtained with {\it HST}. 
\cite{Filippenko1988} highlighted the similarity between the broad H${\alpha}$ and H${\beta}$ emission line profile shapes whereas \citet{Ho1996} illustrated them to be quite different. The new {\it HST}/STIS observations presented in Figure 7 reveals them to be approximately symmetric when wavelength is converted into rest-frame velocity using the non-relativistic Doppler equation. The ratio of the two lines is computed by first smoothing the broad H${\alpha}$ line to the same velocity resolution as H${\beta}$, and then interpolating the H${\alpha}$ flux to the same velocities at
which the H${\beta}$ flux was measured. Consequently, the ratio shares the same velocity resolution as H${\beta}$. The unweighted mean ratio H${\alpha}$/H${\beta}$ = 5.3 ${\pm}$ 1.6 is in good agreement with prior measurements \citep{Peimbert1981, Filippenko1988, Ho1996, Bower1996}. The intrinsic ratio expected for photoionized gas has a value of  2.7 \citep{Osterbrock1989}. By comparison with the much higher observed ratio, one infers a differential extinction between H${\alpha}$ and H${\beta}$ of 0.7 mag. Interestingly, there is some evidence for asymmetric dust extinction {\it internal} to the H$^{+}$ region because the observed ratio systematically exceeds the mean on the blueshifted side, whereas the opposite is true
on the redshifted side, with the exception of three points defining a spike at ${\sim}$ 10$^{3}$ km s$^{-1}$ that is difficult to interpret as it coincides with what may be an incompletely subtracted {[}\ion{N}{ii}{]} line. Of higher statistical significance is a dip in the ratio near zero velocity. Collectively, these trends imply that the H$^{+}$ region suffers internal dust extinction that obscures the inner region more than the outer and obscures the blueshifted side more than the redshifted one.

\subsection{Absolute Extinction to the Central Continuum Source}

Much of what is believed to be understood about the nature of the central UV--X-ray source in M81 was anticipated by \cite{Petre1993a}, including the absence of a geometrically thin accretion disc \citep{Young2018}, consistent with the absence of Compton reflection \citep{Ishisaki1996} and a low accretion rate modulated by an advection dominated accretion flow \citep[ADAF][]{Narayan1994, Narayan1995, Narayan1995a} that is expected to produce a substantial H ionizing continuum \citep{Nemmen2014}\footnotemark \footnotetext{https://figshare.com/articles/Spectral\_models\_for\_low-luminosity\_active\_galactic\_nuclei\_in\_LINERs/4059945}, illustrated in Figure 13 in the context of the observed STIS spectra.
\begin{figure}
	\includegraphics[width=\columnwidth]{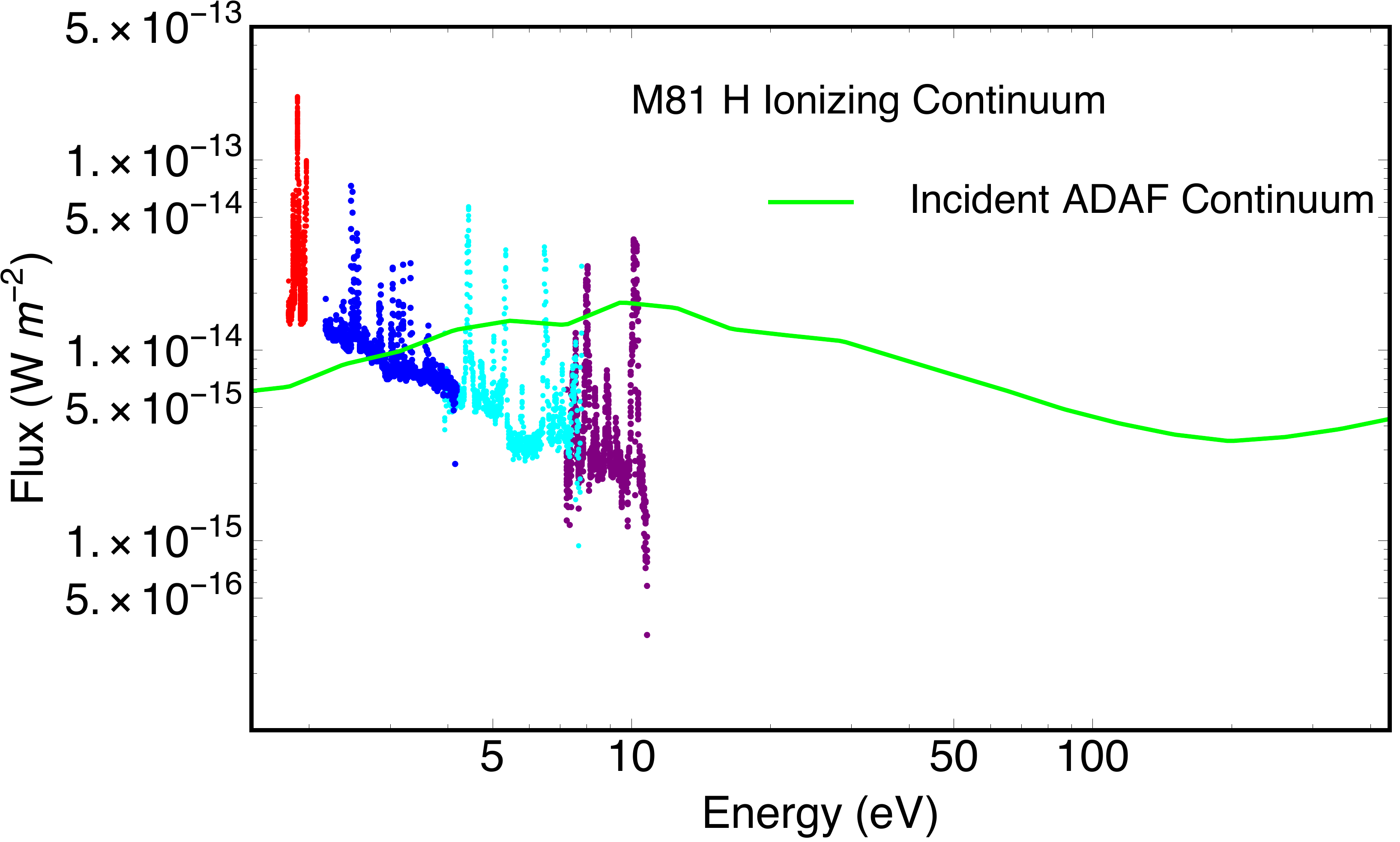}
    \caption{Observed UV--visible continuum of M81 defined by contemporaneous STIS observations with the G750M, G430L, G230L and G140L gratings, depicted by red, blue, indigo and purple dots, respectively. The solid green line represents the incident ADAF continuum of \citet{Nemmen2014}. Units of the ordinate are W/m$^2$ and the abscissa is eV.}
    \label{fig:example_figure}
\end{figure}
Starlight causes the observed visible continuum to be brighter than the incident ADAF continuum. Conversely, dust extinction causes the observed UV continuum to be both fainter and redder than the ADAF. Consequently, one can compare the ADAF continuum with the observed line-free visible--UV continuum to learn about dust extinction and the contribution of starlight, both as a function of wavelength, along the line-of-sight to the continuum source. 

Foreground dust extinction, ${A_\lambda}$ to the UV continuum has been computed for M81 using a method described previously in \cite{Devereux2018} and the result is illustrated in Figure 14. Since starlight obviously dominates the observed continuum longward of 3000 {\AA}, a least squares fit to just the G140L and G230L line-free continua yields,
\begin{equation}
A_{\lambda} = \frac{(2972 \pm 8)}{\lambda(\textrm\AA)}~~~~\rm{mag}
\end{equation}
predicting A${_v}$ = 0.7 and E(B-V) = 0.14 yielding the ratio A${_v}$ = 0.7/E(B-V) = 5, similar to Galactic values. 
A depression in the G230L continuum at ${\sim}$ 5.6 eV
(Figure 13) coincides with the wavelength expected for the 2175 {\AA} feature \citep{Cardelli1989} identified in Figure 14. 

\begin{figure}
	\includegraphics[width=\columnwidth]{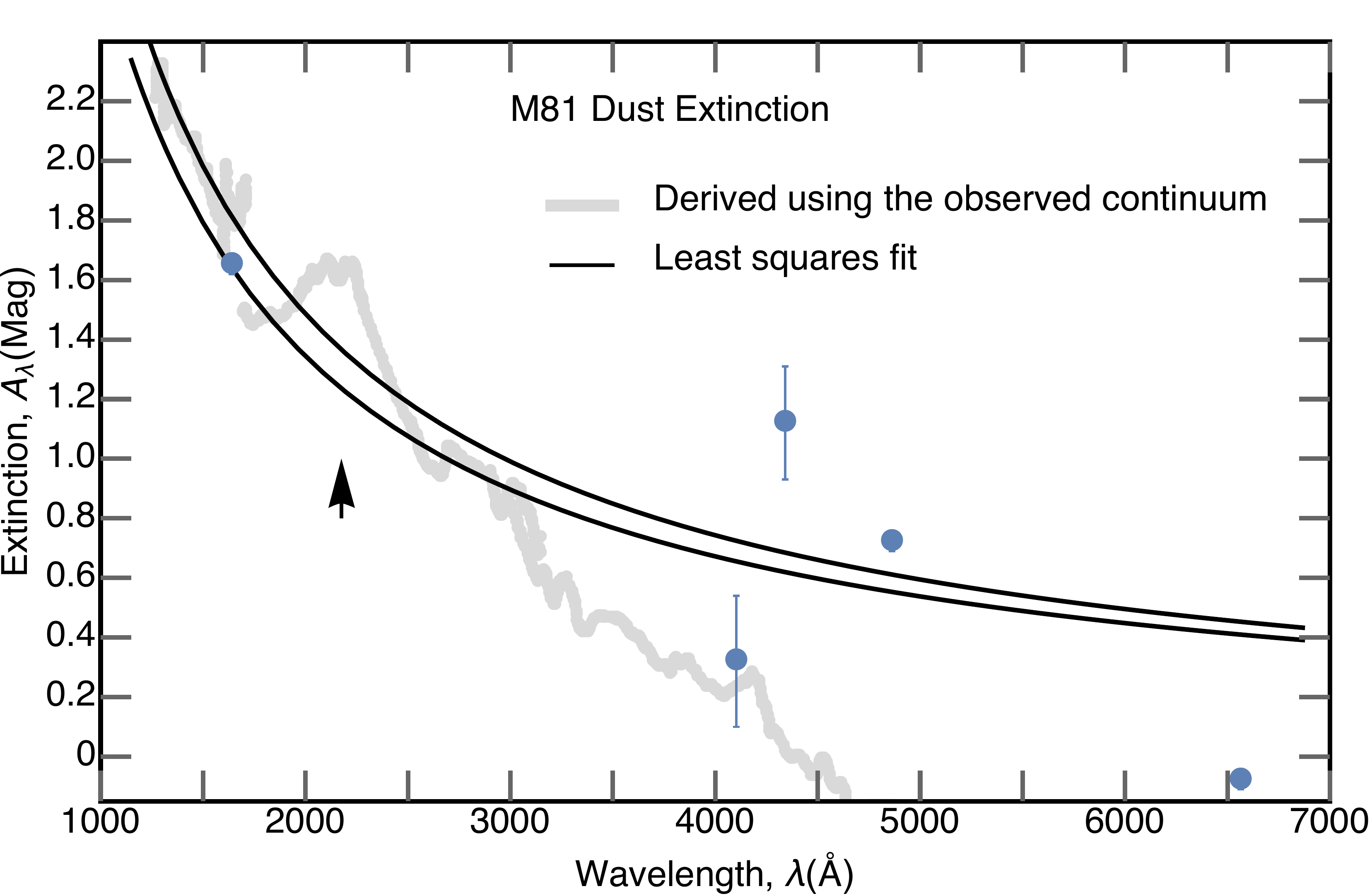}
    \caption{Rest frame extinction to the emission line-free UV--visible continuum in M81 (light grey line) inferred by
comparing contemporaneous STIS observations obtained using the G140L, G230L, G430L and G750M gratings with the intrinsic ADAF continuum of \citet{Nemmen2014}. 
The ordinate indicates total dust extinction in units of magnitudes and the abscissa indicates wavelength in the rest frame of M81 expressed in units of {\AA}. The black arrow identifies the wavelength expected for the 2175 {\AA} feature.
Comparing contemporaneous STIS observations of the H${\alpha}$, H${\beta}$, H${\gamma}$, H${\delta}$ and ${\lambda}$1640 ${\ion{He}{ii}}$ emission lines with the intrinsic emission line fluxes predicted by {\scriptsize XSTAR} defines the extinction to the H${^+}$ region (blue dots, see Table 3). The black curves illustrate the functions described by Equations 1 and 3. }

   \label{fig:example_figure}
\end{figure}

\subsection{Photoionization Modelling of the H$^{+}$ region }

Integrating the ADAF dominated continuum, illustrated in Figure 13, between 1 and 1000 Ryd leads to a luminosity of 7.14 ${\times}$ 10$^{40}$ erg s$^{-1}$, yielding 6.70 ${\times}$ 10$^{50}$ H ionizing photons s$^{-1}$, effectively alleviating the ionizing deficit for the broad Balmer lines reported previously \citep{Ho1996, Bruzual1982, Devereux2007}. The reason being that the ADAF continuum adopted here
is different from the power laws that were employed previously to interpolate
the unobservable ionizing continuum. 
Consequently, a grid of {\scriptsize XSTAR} photoionization models 
\citep{Kallman2001} has been generated to explore the possibility that the H$^{+}$ region responsible for producing the broad H${\alpha}$ emission line is photoionized by the central UV--X-ray source. The methodology 
involves using the shape and luminosity of the bright and broad H${\alpha}$ emission line to constrain a unique photoionization model. The utility of this approach has been demonstrated previously \citep{Devereux2018}. 

The  {\scriptsize XSTAR} models invoke a sphere with a covering factor and a filling factor both equal to unity, and an ionization parameter that is factors of 10${^2}$ -- 10${^3}$ higher than considered for dense broad-line clouds by \citet{Ho1996}. The models are neither isobaric nor isothermal and no column density restriction was imposed. Each {\scriptsize XSTAR} model produces a radial distribution for the H${\alpha}$ emissivity that is surprisingly sensitive to the input radial distribution of neutral gas described by a radial particle number density ${\rho}(r)$ and a power law of index $n$, normalized by a number density ${\rho}_o$, at a reference radius $r_o$, so that ${\rho}(r)$=${\rho}_o(r/r_o)^{-n}$, where ${r}$ is the radial distance from the photoionizing source. Such models produce a wide variety of H${\alpha}$ emission line profile shapes given a kinematic description for the H$^{+}$ gas. 
Radiation pressure is negligible since the active nucleus radiates at 3 x 10${^{-5}}$ of the Eddington luminosity limit \citep{Nemmen2014}.  Thus, the dominant force is gravity implying velocity laws of the form ${v(r)}$ = ${\sqrt{ \gamma G M(r)/r}}$, where ${v}$ is velocity, ${\gamma}$ is a constant that depends on geometry, ${G}$ is the gravitational constant, ${M(r)}$ is the mass interior to ${r}$, where ${r}$ is the radial distance of each point from the central supermassive black hole (BH). The radial dependence of the central mass arises because ${M(r)}$ includes both the BH mass ${M_{\bullet}}$, and the surrounding stars ${M_{\star}(r)}$, both of which are specified in \cite{Devereux2003}.
The H$^{+}$ gas is unlikely to be rotating in a geometrically thin accretion disc, for which ${\gamma}$ = 1, as there does not appear to be one \citep{Young2018}.
Besides, such a disc would produce a double-peaked and possibly asymmetric H${\alpha}$ emission line profile, unless the disc is contrived to be nearly face-on  \citep{Chen1989}. Points moving radially at the escape velocity, for which ${\gamma}$ = 2, is equivalent to those same points on randomly oriented circular orbits, except the velocity amplitude will be smaller for the latter by a factor of ${\sqrt{2}}$. 
Radial motion readily produces symmetric, single-peak, profiles like the observed one, an example of which is illustrated in Figure 15. Although kinematics alone cannot distinguish between various geometries, photoionization modelling provides an important 
additional geometric constraint in the form of a covering factor for the line-emitting gas. Thus, the best way to explore the geometry of the line-emitting gas is to combine the
two methods.

\begin{figure}
	\includegraphics[width=\columnwidth]{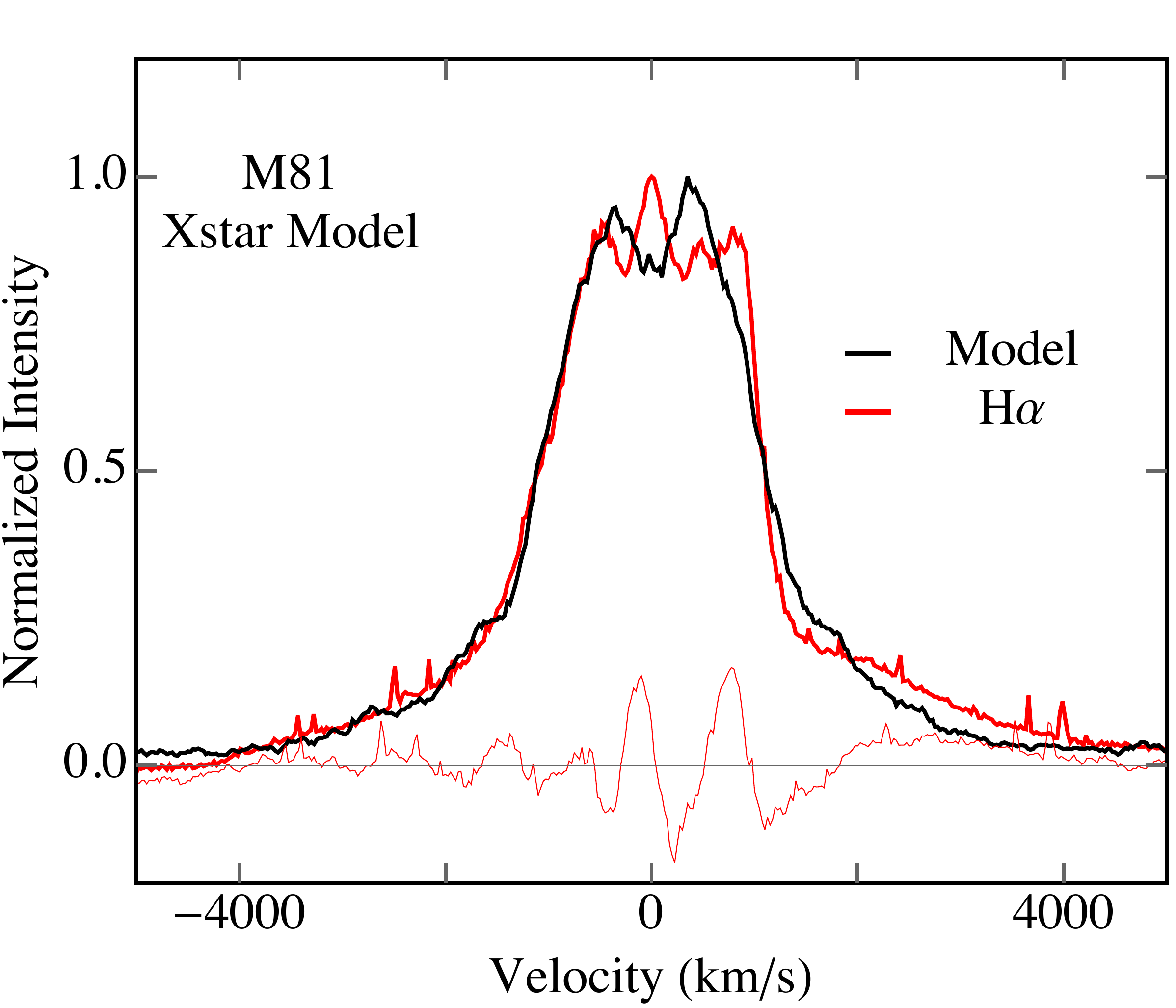}
    \caption{The observed normalized H${\alpha}$ emission line profile is depicted by the red line. The normalized model line, shown in black, corresponds 
to the location of ${\chi}_{red}^2$ minimum in Figure 17 where ${\xi}$ = 3.25 dex at $r_o$ = 5 ${\times}$ 10$^{-3}$ pc, for n = 0.65. The residual between the 
observed and model line is represented by the thinner red line. The abscissa indicates rest frame radial velocity of the H${\alpha}$ emission in km s$^{-1}$ (see Section 3.6 for details).}
    \label{fig:example_figure}
\end{figure}

In a procedure that amounts to an exploration of parameter space, the observed H${\alpha}$ emission line profile shape is compared with simulated ones 
generated by a grid of photoionization models
defined by three parameters; the ionization parameter ${\xi}$, which together with a number density ${\rho}_o$ 
defines an inner radius $r_o$, and a power law of index $n$, which describes the degree of central concentration 
for the spherical ball of gas that is to be photoionized.
The reduced chi-squared statistic
\begin{equation}
{\chi}_{red}^2 = {\sum_j} (O_j - M_j)^2/(\nu \delta^2)
\end{equation}
is used to compare the observed normalized line profile intensities,  ${O_j}$ with the model ones, ${M_j}$. There are 471 degrees of freedom, ${\nu}$,
and the uncertainty, ${\delta}$, in the observed normalized line profile intensities is 4\%. The summation was performed over the velocity span of the broad H${\alpha}$ line depicted in Figure 15. The observed profile has a flat top that can be reproduced in the model line profiles by creating a bi-conical-shaped cavity that is expected to be occupied by the jets, evidence for which has been seen in the radio \citep{Marti-Vidal2011} and in the visible \citep{Ricci2015b}. At the expense of one more parameter, a bi-conical cavity is invoked in spherical coordinates by restricting the polar angle, ${\theta}$ such that -0.85${\pi}$/2~${\le}~{\theta}~{\le}~0.85{\pi}$/2 for all azimuthal angles ${\phi}$ so that 0 ${\le}$ ${\phi}$ ${\le}$ 2${\pi}$.  This particular opening angle was constrained by an empirical iterative process that minimized ${\chi}_{red}^2$.  
It is intermediate 
between that measured for the precessing radio jet by \cite{Marti-Vidal2011} and depicted in the principal component analysis of \citet[][see their Figure 2]{Ricci2015b}. The resulting geometry for the H${\alpha}$ emitting region is illustrated in Figure 16 wherein the azimuthal axis of symmetry is perpendicular to the line-of-sight. Observationally, M81* has what appears to be a one-sided radio jet, the inclination angle of which is undetermined \citep{Marti-Vidal2011,Bietenholz2000, Bietenholz1996}, but it is quite likely to be small as the one-sided nature of the radio-jet has been attributed to Doppler boosting by \cite{Ricci2015b} based on a 
re-examination of the \cite{SchnorrMueller2011} dataset. More complicated models for the H${\alpha}$ emitting geometry could be contrived, but at the expense of additional free parameters. A merit of the {\scriptsize XSTAR} photoionization models is that the inner and outer radii of the H$^{+}$ region are determined by the Balmer emissivity. Consequently, they do not add to the 4 free parameters already described.

\begin{figure}
	\includegraphics[width=\columnwidth]{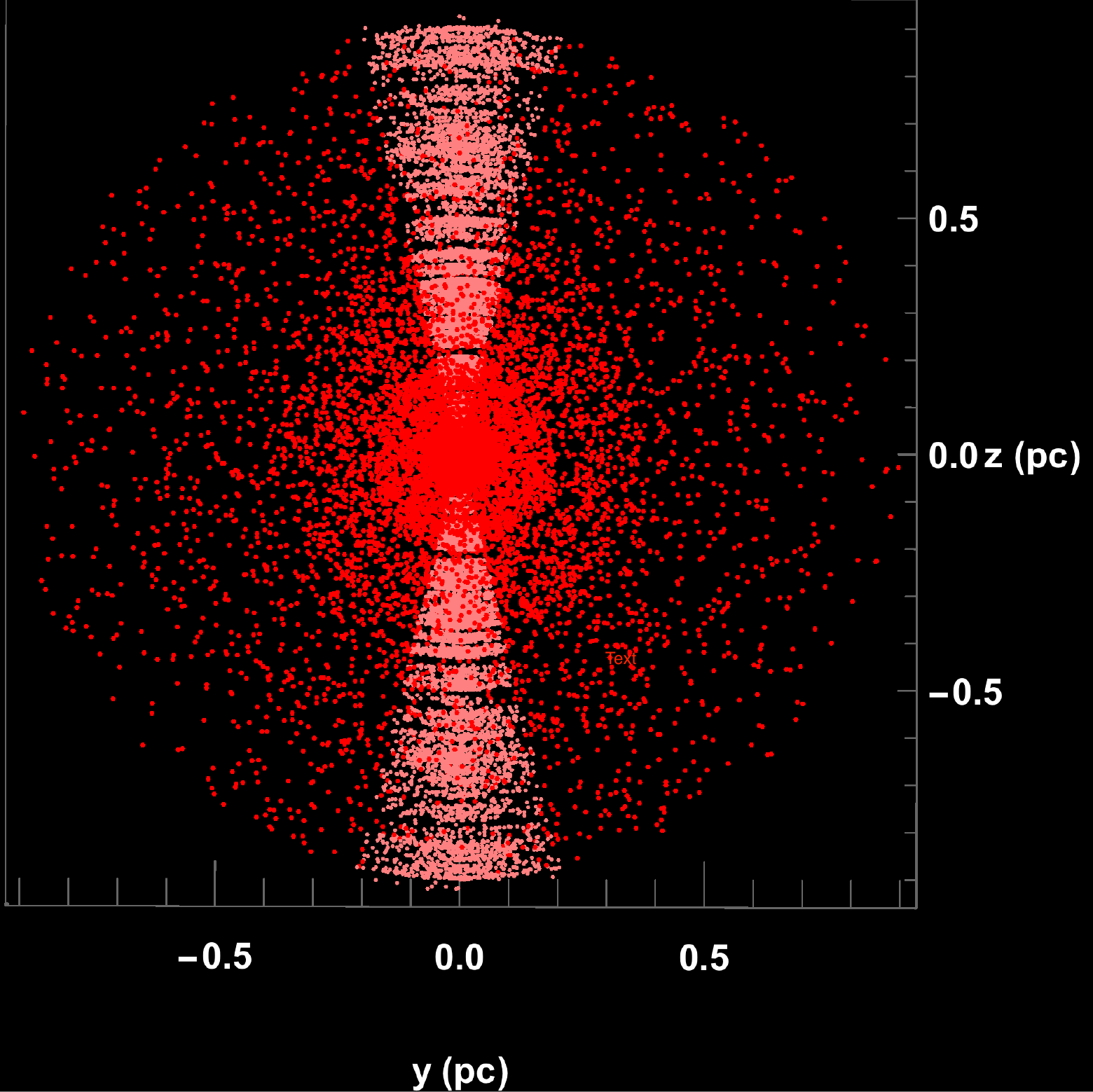}
    \caption{Geometry for the region producing the model H${\alpha}$ emission line profile shown in Figure 15. Dark red dots depict the H${\alpha}$ emissivity.
Lighter pink dots illustrate the jet cavity (see Section 3.6 for details). The ordinate and abscissa are in units of pc.}
    \label{fig:example_figure}
\end{figure}

The default {\scriptsize XSTAR}\footnotemark
\footnotetext{https://heasarc.gsfc.nasa.gov/xstar/docs/html/xstarmanual.html} solar abundance is presumed for H and He. The abundances for other elements are highly uncertain. However, the ${[}\ion{O}{iii}{]}$${\lambda}$5008 emission line is similar in brightness to H${\beta}$ (Figure 3). Prior photoionization models have shown the 
reddening insensitive ratio ${[}\ion{O}{iii}{]}$5008/H${\beta}$ to be a metallicity indicator when the ionization parameter is held constant \citep{Komossa1997, Binette1985, Ferland1983, Halpern1983}. M81 has much in common with NGC 3998  for which 
{\scriptsize XSTAR} models indicate that ${[}\ion{O}{iii}{]}$5008/H${\beta}$ = 60 $Z/Z_\odot$ for 10$^{-3}$ ${\le}$ $Z/Z_\odot$ ${\le}$ 10$^{-1}$ \citep{Devereux2018}.
Thus, for {\scriptsize XSTAR} to correctly predict the observed reddening insensitive ratio ${[}\ion{O}{iii}{]}$5008/H${\beta}$ = 1 ${\pm}$ 0.1 observed for M81 (Section 3.1) the O abundance has to be 2\% solar (Z${\odot}$). Consequently, a grid of models employed abundances equal to 0.02Z${\odot}$, and, for the purposes of comparison, another grid employed solar abundances. The results for the 0.02Z${\odot}$ grid are presented in Figure 17 in the form of a contour plot that clearly identifies a viable {\scriptsize XSTAR} model of interest defined by the parameters, ${\rho}_o$ = 5.05 dex at r$_o$ = 0.01 pc, where the ionization parameter ${\xi}$ = 2.85 dex, for $n$=0.65. The 1.8 pc diameter determined for the H$^{+}$ region, illustrated in Figure 16, is completely contained within the 0.2 arc sec slit that was employed to observe it. A similar solution, not shown, is obtained for solar abundances, but that model was rejected because it overestimates ${[}\ion{O}{iii}{]}$5008/H${\beta}$ by an order of magnitude.

\begin{figure}
	\includegraphics[width=\columnwidth]{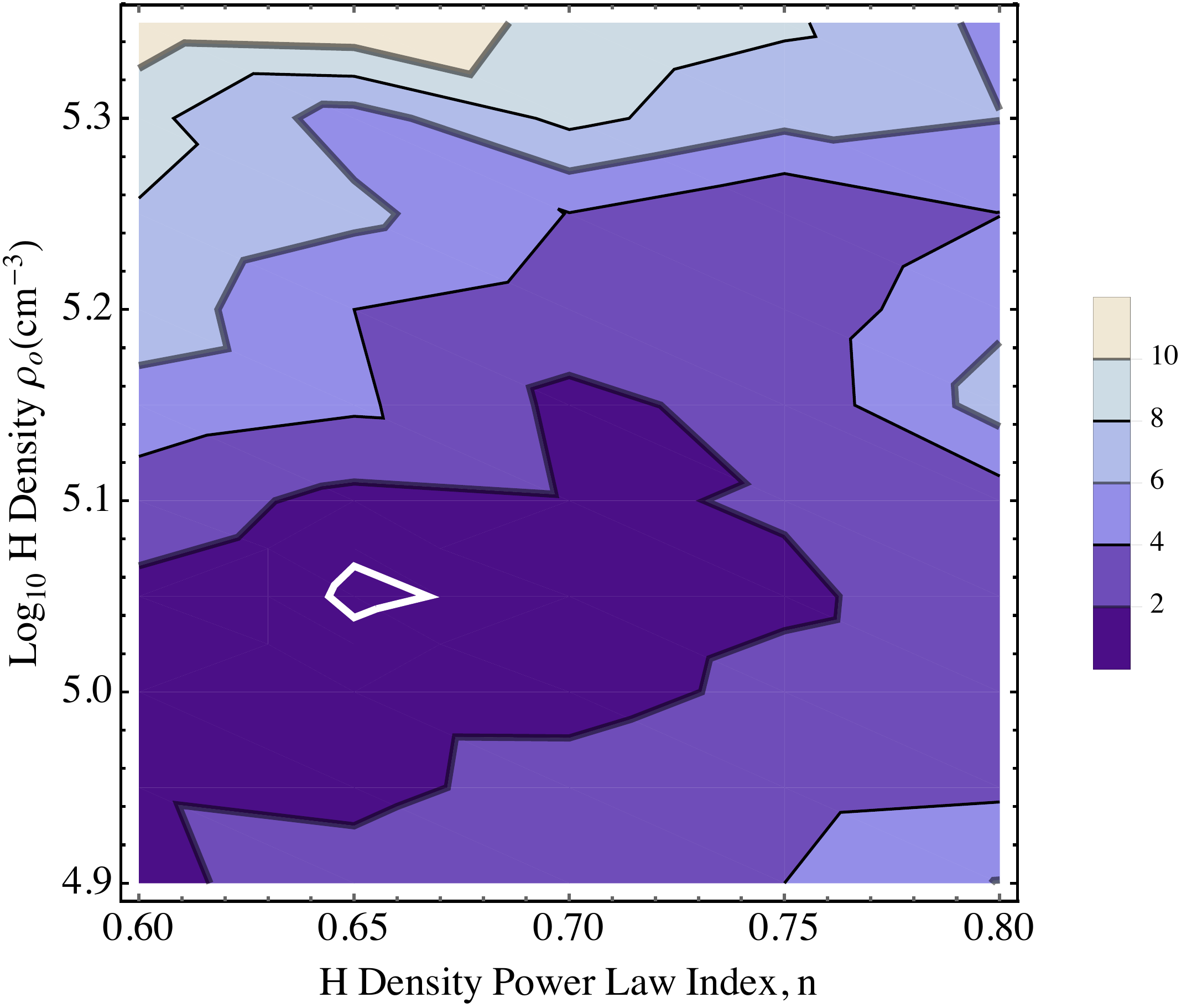}
    \caption{Photoionization model results as a function of gas density ${\rho}_o$ and index $n$ for a metallicity $Z/Z_\odot$ = 0.02. The ordinate identifies the H density ${\rho}_o$ at a reference radius $r_o$ = 0.01 pc. Lines of constant ${\xi}$ run parallel to the abscissa which identifies the index $n$ of the power law used to describe the radial distribution of neutral H gas to be ionized (see Section 3.5 for details). Contours identify ${\chi}_{red}^2$ representing the goodness of fit between the model H${\alpha}$ emission line profile and the observed one (see Figure 15 and Section 3.6)  for M$_{\bullet}$ = 7 ${\times}$ 10${^{7}}$ M${_{\sun}}$. Six contours are plotted. The smallest (white) closed contour represents ${\chi}_{red}^2$ =1.5 and constrains an interesting model with log ${\rho}_o$(cm$^{-3})$ = 5.05, and n = 0.65. The next contour represents ${\chi}_{red}^2$ = 2, and subsequent contours increase in steps of two as indicated by the legend. }
    \label{fig:example_figure}
\end{figure}

\begin{table}
\centering
\caption{Bright H and He recombination lines used to define dust extinction to the H$^{+}$ region}
\begin{tabular}{cccc}
\hline
Line  & Observed & {\scriptsize XSTAR} & Extinction   \\
 & 10$^{38}$ erg s$^{-1}$ & 10$^{38}$ erg s$^{-1}$ & mag \\
(1) & (2) &  (3)  & (4)  \\
\hline
H${\alpha}$ & 15.48 & 14.56 & -0.08 \\
H${\beta}$ &  2.59 & 5.45 & 0.8 \\
H${\gamma}$ & 0.85 & 2.39 & 1.12 \\
H${\delta}$ & 0.50 & 0.67 & 0.32 \\
${\lambda}$1640 ${\ion{He}{ii}}$ & 0.91 & 4.2 & 1.65 \\
\hline
\end{tabular}
\end{table}

Dust extinction to the H$^{+}$ region is determined by comparing the observed H and He recombination line luminosities with those predicted by {\scriptsize XSTAR}. 
A least squares fit to the extinction reported in Table 3 for the broad H and He emission lines yields 
\begin{equation}
A_{\lambda} = \frac{(2688 \pm 535)}{\lambda(\textrm\AA)}~~~~\rm{mag}
\end{equation}
The larger uncertainty in the extinction law predicted by the recombination lines (Equation 3), compared to the line-free continuum (Equation 1), reflects the smaller number of points involved in the fit. Collectively, the results presented in Figure 14 show that the foreground extinction to the H$^{+}$ region is, statistically, indistinguishable from that to the ADAF continuum and is in agreement with prior estimates \citep{Peimbert1981, Filippenko1988, Ho1996}. According to this particular {\scriptsize XSTAR} photoionization model, the results depicted in Figure 14 indicate that
the H$^{+}$ region is completely unobscured at the wavelength of the H${\alpha}$ line, and most of the ${\sim}$ 0.8 mag of differential extinction between H${\alpha}$ and H${\beta}$ (Table 3) is caused by dust internal to the H$^{+}$ region described previously in Section 3.4.

\section{Discussion}

The new STIS observations of M81 reaffirm prior spectroscopic measurements including the detection of several broad emission lines spanning the visible to the UV, an anomalous H${\alpha}$/H${\beta}$ ratio, an astonishingly low ${[}\ion{O}{iii}{]}$5008/${[}\ion{O}{iii}{]}$4364 ratio, and a broad H${\alpha}$ emission line that is redshifted relative to the narrow-line region. What STIS has revealed that is new are several previously unreported broad forbidden emission lines in the UV that have extended the correlation between FWHM and critical density by 
several orders of magnitude (Figure 6), and a broad \ion{C}{iv} ${\lambda}$1548 line that has doubled in brightness (Figure 12). 
The evidence for a redshifted BLR has diminished (Figure 10), although future sub-arcsecond imaging of the nucleus of M81 will likely reveal emission line images
with slightly different centroids depending on wavelength (Figure 8). 

\begin{figure}
	\includegraphics[width=\columnwidth]{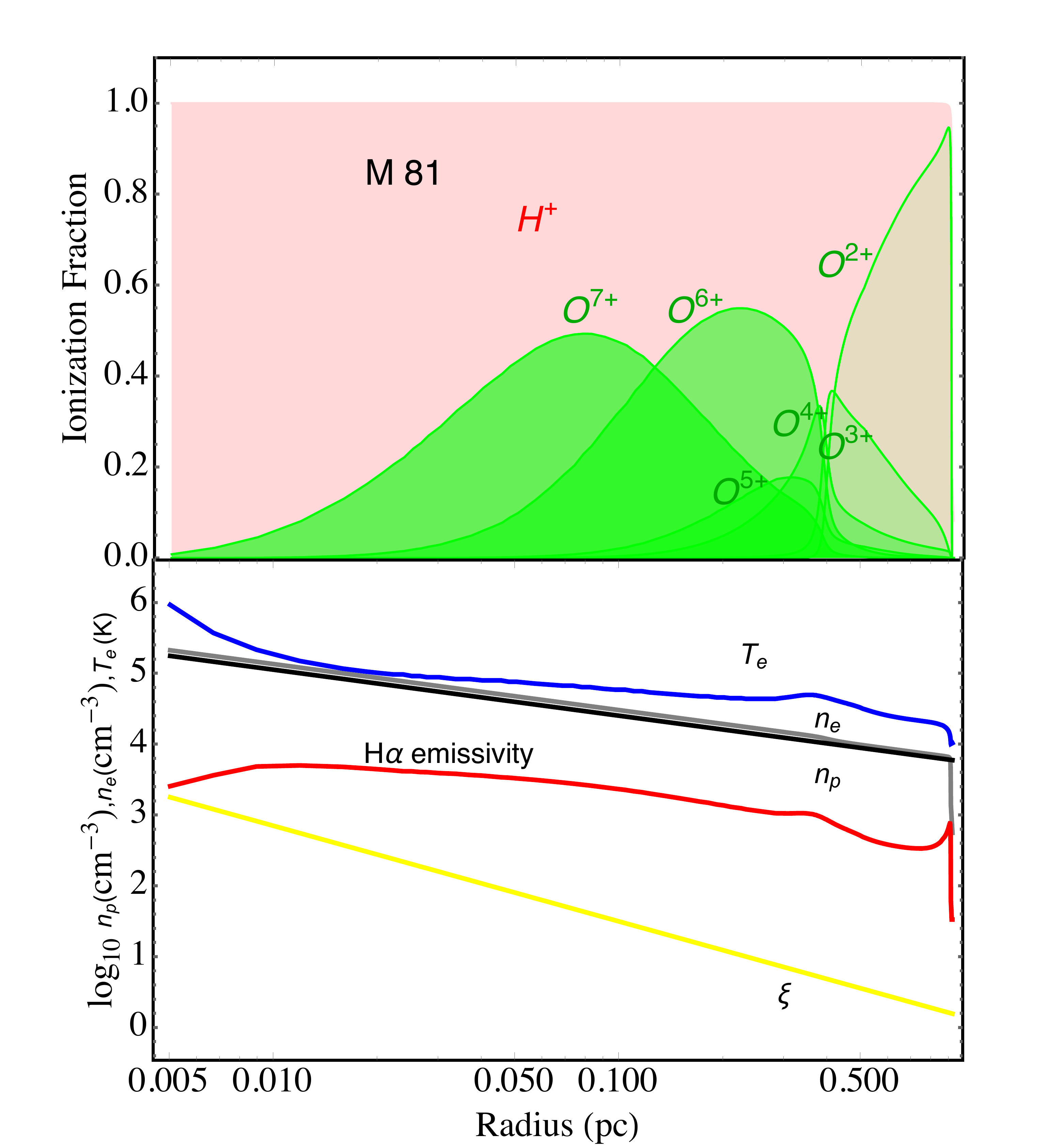}
    \caption{Radial distributions for the photoionization model that best reproduces the observed 
H${\alpha}$ emission line profile shape and luminosity. ${\xi}$ = 3.25 dex at the smallest radius plotted corresponding to 5 ${\times}$ 10$^{-3}$ pc. 
The ordinate refers to a variety of units whereas the abscissa indicates distance from the BH in pc. {\sl Upper panel}: 
Ionization fractions of H${^+}$ (pink shading) and all possible ionization stages of O (green shading, O$^+$ is imperceptible).  {\sl Lower panel}: Log$_{10}$ of
the electron temperature, T$_e$(K)  (blue line), electron density, n$_e$ (cm$^{-3}$) (grey line), proton density, n$_p$ (cm$^{-3}$) (black line), the H${\alpha}$ emissivity in arbitrary units (red line) and the ionization parameter ${\xi}$ (yellow line). }
    \label{fig:example_figure}
\end{figure}

\begin{figure}
	\includegraphics[width=\columnwidth]{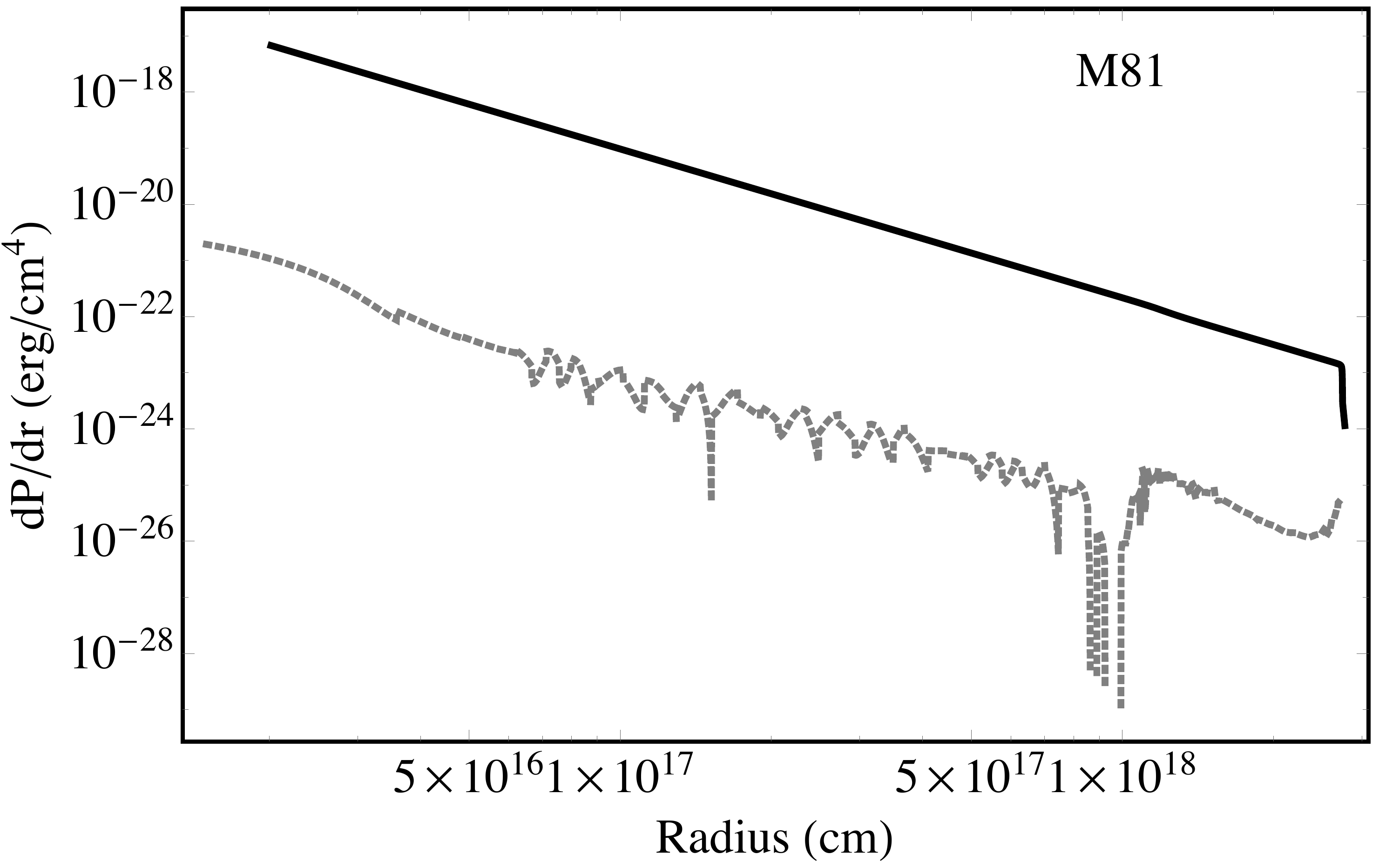}
    \caption{Radial distributions illustrating the gradient in the gravitational energy density (black solid line) and the negative of the gas pressure gradient (gray dotted line) within the H${^+}$ region. The ordinate refers to units of erg cm${^{-4}}$ whereas the abscissa indicates distance from the BH in cm. The figure shows that the H${^+}$ region is not in hydrostatic equilibrium.  }
    \label{fig:example_figure}
\end{figure}

Interpreting the shape of the broad H${\alpha}$ emission line in terms of an H$^{+}$ region, photoionized by the ADAF that is believed to power the 
low-luminosity AGN in M81 \citep{Nemmen2014, Petre1993a}, provides a natural explanation for the size of the region emitting the Balmer lines. Radial distributions are illustrated for several interesting physical parameters in Figure 18. The H$^{+}$ region is ionization bounded with an outer radius of ${\sim}$ 0.9 pc. As the central UV--X-ray source is approached 
the electron temperature increases to ${\sim}$ 10$^{6}$ K transforming the photoionized gas into a plasma, marked by a sharp decline in the Balmer 
emissivity at an inner radius of ${\sim}$ 5 ${\times}$ 10$^{-3}$ pc that is approximately coincident with the transition radius ${r_{tr}}$ = 2 ${\times}$ 10$^{-3}$ pc
parametrized by \citet{Nemmen2014}. Integrating between the inner and outer radii leads to a H equivalent column density of 3.9 ${\times}$ 10$^{22}$ atoms cm$^{-2}$. Integrating the total mass of H between the inner and outer radii results in 532 M${\odot}$, assuming spherical symmetry. Figure 19 shows that within the H$^{+}$ region, the magnitude of the gradient in the gravitational energy density opposes that of the gas pressure by over one order of magnitude. Consequently, mass inflow is inevitable. Assuming spherical symmetry, the corresponding mass inflow rate can be computed using the equation of continuity. Adopting the same unity filling 
factor as employed in the photoionization models (Section 3.6), and a free-fall velocity of 7892 km s$^{-1}$ at a radius of 10$^{-2}$ pc where the proton number 
density is 10${^5}$ cm$^{-3}$, results in $\dot{m}$ = 3 ${\times}$ 10$^{-2}$ M${_{\sun}}$/yr which exceeds, by a factor of 6, the inflow rate required to sustain the ADAF modelled by \cite{Nemmen2014}. The inferred mass inflow rate also exceeds, by a similar factor, that estimated for a much larger region in M81 studied by \cite{SchnorrMueller2011}, $\dot{m}$ also exceeds the minimum expected to power the radio jet \citep{Bietenholz1996}, and the rate estimated for the jet outflow \citep{Nemmen2014}, by three to four orders of magnitude. 

\cite{Filippenko1988} suggested that the anomalous H${\alpha}$/H${\beta}$ ratio observed for the broad emission lines implies high gas density rather than dust extinction. The {\scriptsize XSTAR} photoionization models do produce an anomalous H${\alpha}$/H${\beta}$ ratio, similar to that observed, if the 
density within the H$^{+}$ region is ${\sim}$ 10$^{7}$ cm$^{-3}$, but the resulting small size for the emitting region ${\sim}$ 10$^{-2}$ pc, would produce a very broad H${\alpha}$ emission line, incompatible with the observed one. In fact, the shape of the broad H${\alpha}$ emission line is very sensitive to the radial distribution and
density of the photoionized gas, as Figure 17 illustrates. For a BH mass of 7 ${\times}$ 10$^{7}$ M${_{\sun}}$, the observed H${\alpha}$ emission line profile shape is consistent with a lower density H$^{+}$ region ${\sim}$ 10$^{4}$ to 10$^{5}$ cm$^{-3}$ (Figure 18), in which case the anomalous H${\alpha}$/H${\beta}$ ratio is attributed to dust extinction (Figure 14). Large micron sized dust grains are expected to survive in the H$^{+}$ region because the central UV--X-ray source is too feeble to sublimate them \citep{Barvainis1987}. Furthermore, the existence of dust is evidenced by the reddening reported in Sections 3.4, 3.5 and 3.6. However, there does appear to be considerably less dust than one would expect from the H column which predicts ${\sim}$ 2.5 mag of visual extinction, far more than is actually observed.  
Nevertheless, in the context of an inflow, the velocity dependence of the reddening sensibly indicates that the far-side of the H$^{+}$ region is more obscured than the near-side (Figure 7).

A wide range of gas density, increasing towards the central BH, is implied by the correlation between FWHM and critical density depicted in Figure 6.
The range of densities is much wider than expected to be produced by photoionization of the H gas by the central UV--X-ray source (Figure 18).
The forbidden and semi-forbidden lines are also unlikely to be identified with the atmospheres of stars because the density at the penetration depth of the ionizing photons is too low, ${\leq}$ 10$^{8}$ cm$^{-3}$ \citep{Devereux2007}. The relative brightness of the forbidden and semi-forbidden emission lines does resemble that expected from shocks, in particular, the average of the 200 km s$^{-1}$ and 250 km s$^{-1}$ models (s20010 and s25010) in \cite{Dopita1996}. Shocks could produce the correlation illustrated in Figure 6 if the forbidden and semi-forbidden lines are produced by the interaction of the jet with the surrounding medium, the geometry of which is depicted in Figure 16.

The conundrum with explaining the ${[}\ion{O}{iii}{]}$ emission lines in the context of photoionization is that the observed ${[}\ion{O}{iii}{]}$${\lambda}$5008 line is 
as faint as H${\beta}$, which sets a firm upper 
limit of 2\% solar for the metallicity of the photoionized gas. However,  the remaining discrepancy, that the observed
${[}\ion{O}{iii}{]}$${\lambda}$4364 line is too bright, by about a factor of 3, relative to H${\gamma}$ cannot be 
diminished by further adjusting the metallicity. Thus, the nucleus of M81 has what is known as an ${[}\ion{O}{iii}{]}$${\lambda}$4364 {\it temperature problem} that has afflicted other photoionization models of Liners and Seyfert 2s \citep{Binette1996, Komossa1997, Richardson2014}.  Apparently, the resolution of this problem, in M81 at least, is that the ${[}\ion{O}{iii}{]}$ lines, and, presumably, all the other forbidden and semi-forbidden emission lines, are produced by the interaction of the jet with the surrounding gas
in agreement with the interpretation of \cite{Dopita2015} who noted the similarity between the emission line spectrum of M81 and 
NGC 1052; a Liner with a spatially resolved pair of jets.  However, the main difference between M81 and NGC 1052 is that the former is much closer. 
Thus, the {\it HST} observations 
of M81 probe a much smaller, and unresolved region near the base of its jets, where the gas has been compressed over a much wider range of densities (Figure 6) than imagined by \cite{Dopita2015}. Interaction of the jet with the surrounding medium could also explain the time-variable \ion{C}{iv} ${\lambda}$1548 line emission which has brightened by a factor of 2 over a wide range of radii (Figure 12) even though the adjacent continuum has remained unchanged. However, the jets in NGC 1052 are not a strong source of Balmer emission \citep{Sugai2005, Dopita2015}. Presumably, the same is also true in M81. Consequently, the broad Balmer emission lines
are unlikely to be identified with the jets or jet cavity.

According to {\scriptsize XSTAR} the central ADAF is expected to progressively ionize O through all of its possible ionization stages as the central UV--X-ray source is approached (Figure 18). Interestingly, emission lines from \ion{O}{vii} and \ion{O}{viii} have been reported by \cite{Page2003} and \cite{Young2007} at about
the same brightness as predicted by the 2\% solar metallicity {\scriptsize XSTAR} photoionization model. Thus, the photoionization model presented here for 
the low-luminosity nucleus of M81 may well reconcile recombination lines seen from the visible to the X-rays with a single photoionized region consisting of low metallicity gas.

How did the nucleus of M81 acquire 532 M${\odot}$ of low metallicity gas? The answer may be that a dwarf galaxy fell in. Dwarf galaxies exhibit a fascinating correlation between metallicity and mass in the sense that the lower the mass, the lower the metallicity \citep[][and references therein]{Lee2006}.
Observational evidence that our Galaxy has merged with a dwarf has been presented by \citet{Helmi2018}.  A similar merger could account for the low metallicity, and low dust content of the gas in the nucleus of M81, as well as its velocity dispersion (Figure 8). \citet{Morse1996} proposed this idea, in a similar context, for other low-luminosity AGN. 

\section{Conclusion}

M81 is classified as both a Seyfert 1 and a Liner because it exhibits emission lines of a variety of widths covering a wide range of ionization potential and critical density. 
Collectively, the visible--UV spectrum of M81 appears to be composed of at least three separate emission line components
that are not kinematically distinct. Firstly, condensations of a wide range in density, 10$^{5}$ to 10$^{11}$ cm$^{-3}$, and low covering factor, to explain the collisionally excited forbidden lines.  Secondly,  a lower density, 10$^{4}$ to 10$^{5}$ cm$^{-3}$, photoionized region of high covering factor to explain the Balmer lines. Thirdly, an extended source of time-variable \ion{C}{iv} ${\lambda}$1548 emission. These three emission line components are superimposed on a visible--UV continuum that has not varied significantly since the last measurement obtained with ${\it HST}$ 21 years ago. Collectively, these observations can be understood in the context of a jet interacting with an infalling cloud of gas that is photoionized by the central UV--X-ray source. The photoionized region producing the broad Balmer lines is large, ${\sim}$ 1 pc in radius. It consists of ${\sim}$ 500 M${\odot}$ of low metallicity gas that is dynamically unstable, and prone to infall at a rate that is commensurate with that required to power the ADAF for 10$^{5}$ yrs.

\section*{Acknowledgements}

The author is grateful to the many colleagues that he met during the course of his sabbatical tour of universities in the United States, and the United Kingdom, for their kind hospitality, their curiosity and thought-provoking questions that provided much inspiration 
for the concurrent production of this manuscript. Special thanks, in particular, go to Dr. Steven Willner who, generous of his time, has graciously waded through and commented on prior less coherent versions of this manuscript that greatly improved its presentation. Special thanks also to ERAU Cyber Intelligence \& Security undergraduate student Jessica Wilson for writing a very useful piece of Python code that greatly accelerated the analysis described herein. STSDAS is a product of the Space Telescope Science Institute, which is operated by AURA for NASA. Based on observations made with the NASA/ESA Hubble Space Telescope, obtained from the data archive at the Space Telescope Science Institute. STScI is operated by the Association of Universities for Research in Astronomy, Inc. under NASA contract NAS 5-26555.

%%%%%%%%%%%%%%%%%%%%%%%%%%%%%%%%%%%%%%%%%%%%%%%%%%



\bibliographystyle{mnras}
\bibliography{M81v5} % if your bibtex file is called example.bib

%%%%%%%%%%%%%%%%%%%%%%%%%%%%%%%%%%%%%%%%%%%%%%%%%%

%%%%%%%%%%%%%%%%% APPENDICES %%%%%%%%%%%%%%%%%%%%%

%\appendix


%%%%%%%%%%%%%%%%%%%%%%%%%%%%%%%%%%%%%%%%%%%%%%%%%%

% Don't change these lines
\bsp	% typesetting comment
\label{lastpage}
\end{document}